\documentclass[%
 preprint,
 superscriptaddress,
 amsmath,amssymb,
 aps, 
 floatfix,
]{revtex4-2}

\usepackage{graphicx}
\usepackage{dcolumn}
\usepackage{bm}
\usepackage[version=4]{mhchem}
\usepackage{tikz}
\usepackage{lipsum}
\usepackage{miller}
\usepackage{siunitx}
\usepackage{hyperref}

\begin{document}

\title{Emergent reactance induced by the deformation of a current-driven skyrmion lattice}

\author{Matthew T. Littlehales}
\email{matthew.t.littlehales@durham.ac.uk}
\affiliation{Durham University, Department of Physics, South Road, Durham, DH1 3LE, United Kingdom}
\affiliation{ISIS Neutron and Muon Source, Rutherford Appleton Laboratory, Didcot, OX11 0QX, United Kingdom}

\author{Max T. Birch}
\affiliation{RIKEN Center for Emergent Matter Science (CEMS), Wako, Japan}

\author{Akiko Kikkawa}
\affiliation{RIKEN Center for Emergent Matter Science (CEMS), Wako, Japan}

\author{Yasujiro Taguchi}
\affiliation{RIKEN Center for Emergent Matter Science (CEMS), Wako, Japan}

\author{Diego Alba Venero}
\affiliation{ISIS Neutron and Muon Source, Rutherford Appleton Laboratory, Didcot, OX11 0QX, United Kingdom}

\author{Peter D. Hatton}
\affiliation{Durham University, Department of Physics, South Road, Durham, DH1 3LE, United Kingdom}

\author{Naoto Nagaosa}
\affiliation{RIKEN Center for Emergent Matter Science (CEMS), Wako, Japan}
\affiliation{Fundamental Quantum Science Program, TRIP Headquarters, RIKEN, Wako 351-0198, Japan}

\author{Yoshinori Tokura}
\affiliation{RIKEN Center for Emergent Matter Science (CEMS), Wako, Japan}
\affiliation{Department of Applied Physics, University of Tokyo, Tokyo, Japan}
\affiliation{Tokyo College, University of Tokyo, Tokyo, Japan}

\author{Tomoyuki Yokouchi}
\email{tomoyuki.yokouchi@riken.jp}
\affiliation{RIKEN Center for Emergent Matter Science (CEMS), Wako, Japan}

\date{\today}

\begin{abstract}
The interaction between conduction electrons and spin textures gives rise to remarkable phenomena associated with the Berry phase. The Berry phase acquired by conduction electrons acts as an emergent electromagnetic field, facilitating phenomena analogous to classical electromagnetism, such as the Lorentz force and electromagnetic induction. Magnetic skyrmions, spin vortices with non-trivial topology, serve as a key platform for such studies. For example, non-trivial transport responses are recognized as being induced by the emergent Lorentz force and the emergent electromagnetic induction. Despite remarkable progress in skyrmion physics, emergent reactance, in which the phase of an applied AC current is modified by emergent electromagnetism, has not been thoroughly investigated. Here, we report emergent reactance in the prototypical skyrmion-hosting material, \ce{MnSi}. We observe longitudinal and Hall reactance signals as the skyrmion lattice undergoes creep motion, in which the skyrmions deform while moving. The Hall reactance is attributed to the emergent electric field associated with the inertial translational motion arising from the skyrmion effective mass. In contrast, the longitudinal reactance results from the emergent electric fields generated by the phason and spin-tilting modes excited by their deformation. Our findings shed light on the internal deformation degrees of freedom in skyrmions as a important factor for efficient generation of the emergent electric field.

\end{abstract}

\maketitle

As an electron traverses a non-collinear spin texture with its spin direction adiabatically aligned with the underlying spins, the  wavefunction of the electron acquires an additional quantum mechanical phase, termed Berry's phase \cite{Berry1984}. The Berry phase acts as an effective electromagnetic field, also known as the emergent electromagnetic field, and is responsible for a rich variety of quantum electronic phenomena \cite{Nagaosa2010, Zhang2005, Qi2011, Armitage2018}. 

A pivotal playground for emergent electromagnetic fields is offered by topological spin structures, known as skyrmions. A skyrmion is a particle-like spin texture characterized by a topological winding number $N_{\rm{Sk}} = -1 $ and can form a hexagonal lattice (SkL) as an equilibrium state in non-centrosymmetric and frustrated helimagnets \cite{Muhlbauer2009, Birch2020, Kurumaji2019}. Because of their topologically non-trivial and non-coplanar magnetic structure, skyrmions generate an emergent magnetic field $\bm{b}_\mathrm{em}$, which deflects electron trajectories similarly to the normal Hall effect, resulting in an additional Hall conductivity known as the topological Hall effect (THE) \cite{Jonietz2010}. 

\begin{figure*}
    \includegraphics[width = \textwidth]{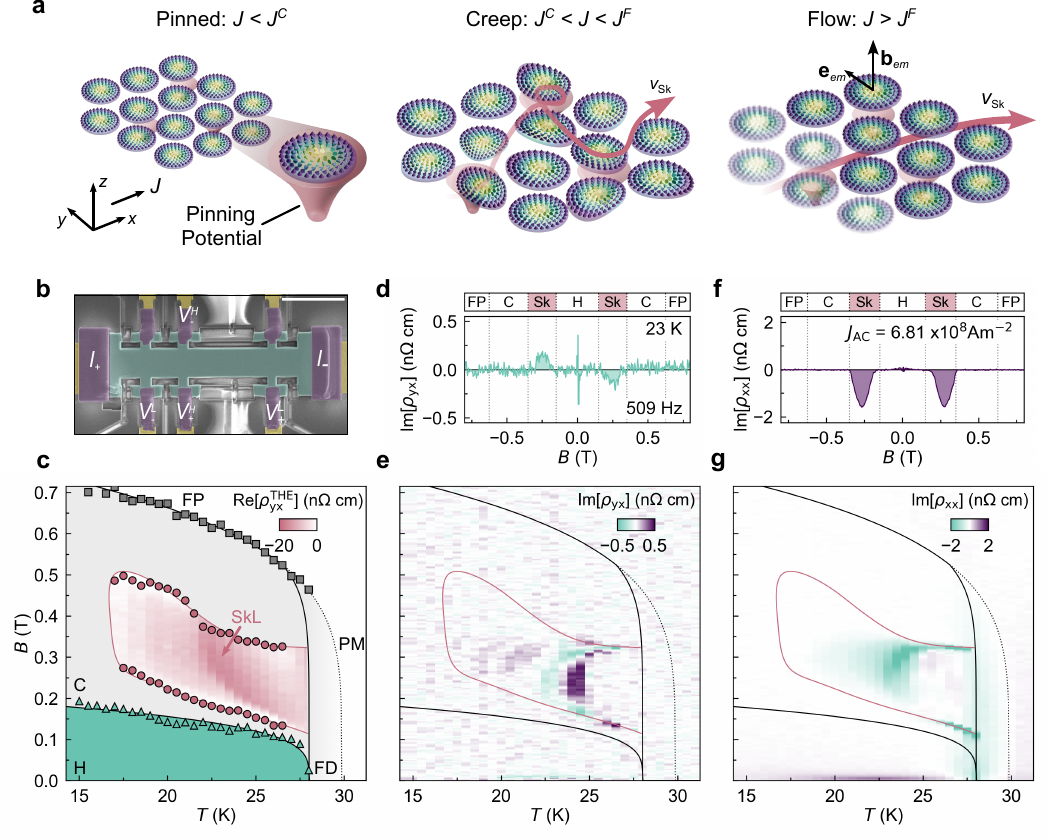}\vspace{-6mm}
    \caption{\textbf{Emergent reactance in the SkL phase of \ce{MnSi}.} \textbf{a} Schematic of the dynamic SkL velocity transition with increasing current density. \textbf{b} Scanning electron micrograph of our micro-fabricated \ce{MnSi} device (Scale bar, \SI{10}{\micro\meter}), labelled with current electrodes $I_\pm$, longitudinal $V_\pm^\mathrm{L}$ and Hall $V_\pm^\mathrm{H}$ voltage terminals. \textbf{c} Magnetic phase diagram of \ce{MnSi} device with helical (H), conical (C), skyrmion lattice (SkL), field polarised (FP), fluctuation disordered (FD), and paramagnetic (PM) phases determined from the Hall resistivity $\rho_\mathrm{yx}$. The SkL phase is overlaid with a colour map of $\rho_\mathrm{yx}^\mathrm{THE}$, obtained by fitting a straight line to $\rho_\mathrm{yx}$ in the conical phase. A detailed procedure of phase boundaries determination can be found in the Methods.  \textbf{d,e} Magnetic field dependence at \SI{23}{\kelvin} (\textbf{d}) and colour map on the magnetic field-temperature plane (\textbf{e}) of the first-harmonic transverse reactance Im[$\rho_\mathrm{yx}$]. \textbf{f,g} Magnetic field dependence at \SI{23}{\kelvin} (\textbf{f}) and colour map on the magnetic field-temperature plane (\textbf{g}) of the longitudinal reactance Im[$\rho_\mathrm{xx}$]. The phase diagram in \textbf{c} was determined from the Hall resistivity measured with a sinusoidal current of amplitude $J_\mathrm{AC} = $ \SI{0.68e8}{\ampere\per\meter\squared}, and Im[$\rho_\mathrm{yx}$] (\textbf{e}) and Im[$\rho_\mathrm{xx}$] (\textbf{g}) with $J_\mathrm{AC} = $ \SI{6.81e8}{\ampere\per\meter\squared}, both at a frequency of $f = $ \SI{509}{\hertz}. The feature at $B = $ \SI{0}{\tesla} in (\textbf{d}) is a result of the extrinsic reactance from the measurement setup (see Supplementary Information)}
    \label{f1}
\end{figure*}

An emergent electric field is associated with the dynamics of the SkL. When an electric current is applied to the SkL, spin-transfer torque induces its motion, leading to dynamic transitions \cite{Peng2021, Luo2020, Birch2024}. At low current densities, the SkL is trapped by pinning potentials. Once the current density $J$ exceeds the pinning threshold $J^\mathrm{C}$ the SkL undergoes creep motion, where skyrmions hop between adjacent pinning sites accompanied by their internal deformation. Notably, this deformation manifests as an effective skyrmion mass, leading to inertial motion \cite{Buttner2015, Birch2024}. Finally, when $J$ surpasses the flow threshold $J^\mathrm{F}$, the SkL flows freely through pinning centres, reducing its deformation and dynamically reordering due to the relative weakening of the pinning force \cite{Reichhardt2015, Yokouchi2017}. Schematically, this is demonstrated in Fig. \ref{f1}a. An emergent electric field can be generally described by\begin{equation}
    e_{\mathrm{em},i} = \frac{h}{2\pi e}\bm{n}\cdot(\partial_i\bm{n}\times\partial_t\bm{n}),
\end{equation}
where $\bm{n}$ is a unit vector parallel to the direction of spins \cite{Schulz2012, Volovik1987, Nagaosa2013}. For the translational motion of the SkL with velocity $\bm{v}_\mathrm{Sk}$, Eq. (1) reduces to $\bm{e}_\mathrm{em} = -\bm{v}_\mathrm{Sk}\times\bm{b}_\mathrm{em}$. This emergent electric field opposes the THE, which is experimentally observed as a reduction in the topological Hall resistivity \cite{Schulz2012, Birch2024}. In particular, in the clean limit of the flow regime, the SkL may catch up to the conduction electron velocity, leading to a complete cancellation of the topological Hall resistivity by the emergent electric field due to emergent Galilean relativity \cite{Zang2011, Schulz2012, Birch2024}. 

Another recently discovered platform for exploring emergent electric fields is the current-induced deformation of non-collinear spin textures such as domain walls and helices. A non-collinear spin texture driven by an electric current undergoes spatial and temporal deformation. This deformation can be described by spin-tilting and phason modes, whose excitation notably induces an out-of-phase emergent electric field relative to the input AC current, equivalent to a reactance \cite{Nagaosa2019, Yokouchi2020}. Thus, this emergent electric field contrasts with that induced by the translational motion of the SkL, which induces an in-phase Hall emergent electric field. In particular, the modulation of the input current phase plays a crucial role in modern AC circuits, which is traditionally governed by capacitors and inductors operating based on classical electromagnetism. Thus, the emergent reactance provides novel functionalities based on emergent electromagnetism, expanding the potential for applications \cite{Nagaosa2019}. 

So far, emergent reactance has been demonstrated in a number of helimagnets and domain wall systems \cite{Kitaori2021, Zhang2023, Kitaori2024} and has been further investigated theoretically \cite{Kurebayashi2021, Ieda2021, Yamane2022, Shimizu2023, Araki2023, Oh2024}. However, an emergent reactance generated by the SkL has not been well explored experimentally \cite{Furuta2023, Scheuchenpflug2025}. In this paper, we report the observation of an emergent reactance generated by the deformation of the SkL under creep motion. Notably, since the threshold current density of the SkL is lower than that of helices and ferromagnetic domain walls \cite{Yamanouchi2004, Kimoto2025}, our findings offer an advantage for potential applications.  

For this study, we chose the prototypical skyrmion host MnSi  \cite{Muhlbauer2009, Neubauer2009, Jonietz2010, Schulz2012}. We fabricated a microscale thin-plate of \ce{MnSi} by focused-ion beam (FIB), and mounted it onto a \ce{CaF2} substrate (Fig. \ref{f1}b), to apply high current densities and enhance the signal-to-noise ratio (see Methods for details). The complex impedance is measured using standard lock-in techniques. In Fig. \ref{f1}c we show the magnetic phase diagram of our device together with a colour map of the topological Hall resistivity in the SkL phase. In addition to the enhancement of two-dimensionality, the compressive strain induced by the differential thermal expansion between the \ce{CaF2} substrate and the \ce{MnSi} thin plate stabilizes the skyrmion phase over an extended temperature region compared with bulk samples \cite{Sato2022}. First, in Figs. \ref{f1}d and f, we present the magnetic field dependence of the transverse and longitudinal reactances, (Im[$\rho_\mathrm{yx}$] and Im[$\rho_\mathrm{xx}$]) measured at \SI{23}{\kelvin} using a high AC current density $J_\mathrm{AC} = $ \SI{6.81e8}{\ampere\per\meter\squared} at \SI{509}{\hertz}. Clearly, non-zero signal is observed only within the SkL phase. Furthermore, $B-T$ colour maps presented in Figs. \ref{f1}e and g also highlight non-zero reactance signals only within the SkL phase.
\begin{figure*}
	\includegraphics{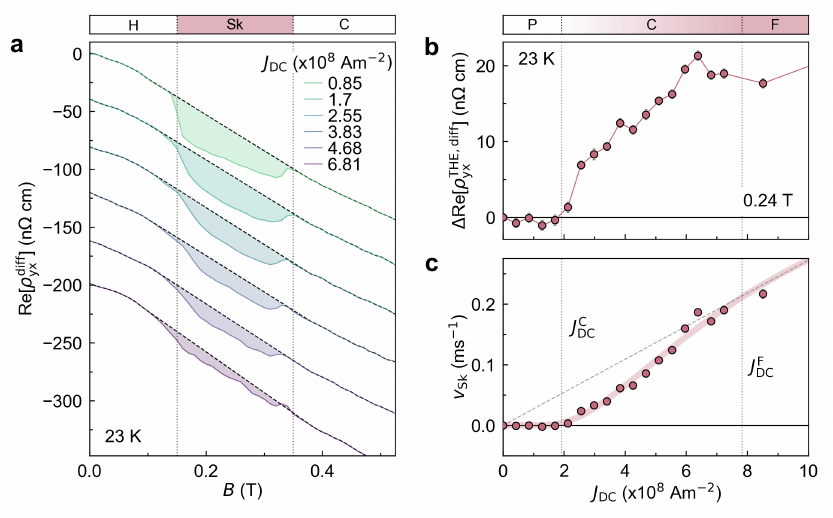}
	\caption{\textbf{Emergent electrodynamics and velocity of the current-induced SkL motion.} \textbf{a} Magnetic field dependence of the differential Hall resistivity Re[$\rho_\mathrm{yx}^\mathrm{diff}$] for various  DC current bias $J_\mathrm{DC}$. The dashed lines denote a spline fit to the regions outside of the SkL phase (see Supplementary Information). \textbf{b} Change in the topological Hall resistivity ($\Delta$Re[$\rho_\mathrm{yx}^\mathrm{THE,diff}$] = Re$[\rho_\mathrm{yx}^\mathrm{THE,diff}](J_\mathrm{DC})$ - Re$[\rho_\mathrm{yx}^\mathrm{THE,diff}]$($J_\mathrm{DC}$ = \SI{0}{\ampere\per\meter\squared}), calculated by subtracting a fitted spline to the regions outside of the topological phase) in the SkL phase at \SI{0.24}{\tesla}. \textbf{c} $\Delta$Re[$\rho_\mathrm{yx}^\mathrm{THE,diff}$] converted into skyrmion velocity $v_\mathrm{Sk}$ using Eq. \ref{velocity}. We plot a red line as a guide to the eye. The dashed line corresponds to 0.75$v_e$ (see Methods for calculation). The dynamic transition of the SkL velocity curve is clearly observed in \textbf{c} with dotted lines indicating the transitions from pinned-creep and creep-flow in each of plots \textbf{b,c}. These measurements were performed by locking in to a small oscillation of $J_\mathrm{AC} = $ \SI{1.74e7}{\ampere\per\meter\squared} at a frequency of $f = $ \SI{509}{\hertz}, modulated on a DC bias, $J_\mathrm{DC}$. }
	\label{f2}
\end{figure*}

 To elucidate the mechanism responsible for the longitudinal and transverse reactance signals, we first determine the dynamical phases of the SkL from the current-density dependence of the topological Hall resistivity. In Fig \ref{f2}a, we present the magnetic field dependence of the differential Hall resistivity Re[$\rho^\mathrm{diff}_\mathrm{yx}$] for selected DC bias current $J_\mathrm{DC}$. Here, the DC bias current drives the SkL, and a superimposed small AC current is used to measure Re[$\rho^\mathrm{diff}_\mathrm{yx}$] (see Methods for details). As $J_\mathrm{DC}$ increases, the topological Hall resistivity (the shaded regions) decreases. This is further highlighted in Fig. \ref{f2}b, where the $J_\mathrm{DC}$ dependence of the change in the topological Hall resistivity $\Delta$Re[$\rho^\mathrm{THE, diff}_\mathrm{yx}$] is plotted for the SkL phase, and displays non-monotonic behaviour closely resembling that observed in bulk \ce{MnSi} \cite{Schulz2012}. The emergent electric field generated by a moving SkL is proportional to its velocity,  which can be expressed  as:
\begin{equation}
    v_\mathrm{Sk} = \frac{J\Delta\rho_{\mathrm{yx}}(J)}{Pb_{\mathrm{em}}},
    \label{velocity}
\end{equation}
in which the SkL velocity clearly reveals itself in a change in the Hall resistivity $\Delta\rho_{\mathrm{yx}}(J)$ \cite{Birch2024}. Using  Eq. (2), we estimate the SkL velocity from $\Delta$Re[$\rho_\mathrm{yx}^\mathrm{THE,diff}$], where $Pb_\mathrm{em}$ is the effective emergent magnetic field and is estimated to be \SI{-0.79}{\tesla} (see Supplementary Fig. S2). Figure \ref{f2}c presents the current density dependence of the estimated SkL velocity, highlighting three distinct regions: (1) $v_\mathrm{Sk} = 0 $ below $J^\mathrm{C}_\mathrm{DC}$; (2) a nonlinear increase in $v_\mathrm{Sk}$ between $J^\mathrm{C}_\mathrm{DC}$ and $J^\mathrm{F}_\mathrm{DC}$; and (3)  $v_\mathrm{Sk}$ proportional to $J_\mathrm{DC}$ and close to the electron velocity $v_e$ above $J^\mathrm{F}_\mathrm{DC}$. Based on previous studies \cite{Luo2020, Peng2021, Birch2024}, we identify the dynamical phases as follows: (1) the pinned state below$J^\mathrm{C}_\mathrm{DC}$, (2) the creep region between $J^\mathrm{C}_\mathrm{DC}$ and $J^\mathrm{F}_\mathrm{DC}$, and (3) the flow regime above  $J^\mathrm{F}_\mathrm{DC}$. We note that  $J_\mathrm{DC}^\mathrm{C}$ and $J_\mathrm{DC}^\mathrm{F}$ in our device are of the same order as those of the microfabricated  sample of \ce{Gd2PdSi3} \cite{Birch2024}, but 2-3 orders of magnitude larger than those in bulk single crystals of MnSi \cite{Schulz2012}. However, the  $v_\mathrm{sk}-J_{\mathrm{DC}}$ profile of our device is qualitatively similar to that in bulk single crystals of \ce{MnSi} \cite{Schulz2012, Luo2020}, and scales with $J_\mathrm{DC}$ nearly identically in the flow region (see Supplementary Information). Accordingly, the higher thresholds can be attributed to enhanced collective pinning, arising from an increased number of pinning sites introduced during the FIB fabrication process and confinement effects in microfabricated samples \cite{Iwasaki2013, Iwasaki2013_2, Birch2024}. Therefore, aside from the higher thresholds, there is no essential difference in the current-driven properties of the SkL between our device and bulk samples.

\begin{figure}
    \includegraphics[width = 0.5\columnwidth]{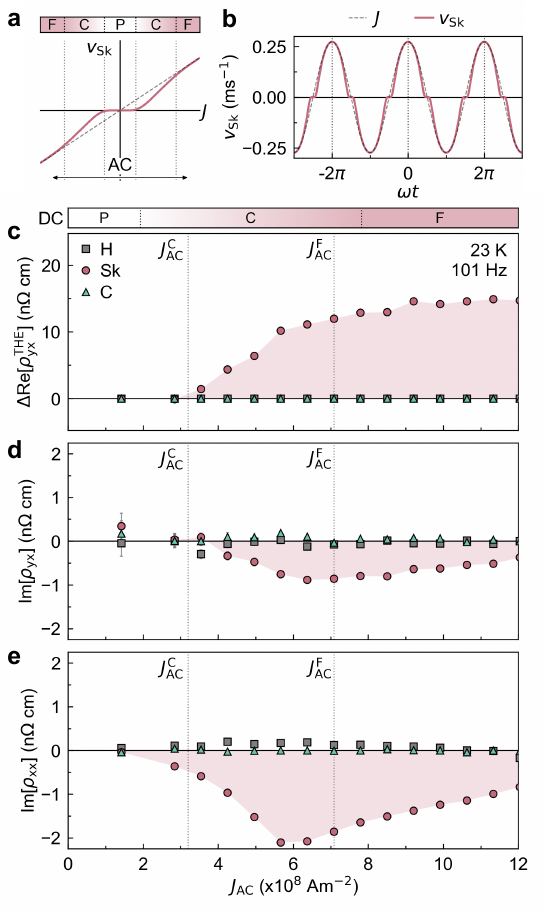}
    \caption{\textbf{Current density dependence of emergent reactance in \ce{MnSi}} \textbf{a,b} Schematic representation of the skyrmion velocity curve for a cycle of AC. The voltage response follows the SkL velocity, but the time-averaged response is primarily governed by the dynamical phase where the current reaches its peak. \textbf{c-e} AC amplitude, $J_\mathrm{AC}$, dependence of $\Delta$Re[$\rho_\mathrm{yx}^\mathrm{THE}]$ (\textbf{c}), Im[$\rho_\mathrm{yx}]$ (\textbf{d}), and Im[$\rho_\mathrm{xx}]$ (\textbf{e}) for helical (H, gray squares), skyrmion lattice (SkL, red circles), and conical phases (C, teal triangles), all measured at \SI{23}{\kelvin} and {\SI{101}{\hertz}}. The threshold current densities for the creep, $J_\mathrm{AC}^\mathrm{C}$, and flow, $J_\mathrm{AC}^\mathrm{F}$, regimes in the low frequency AC regime determined from $\Delta$Re[$\rho_\mathrm{yx}^\mathrm{THE}]$ are indicated in each plot by the dotted vertical lines, and compared with the DC limit (the top panel \textbf{c}.)}
    \label{f3}
\end{figure}

\begin{figure*}
    \includegraphics{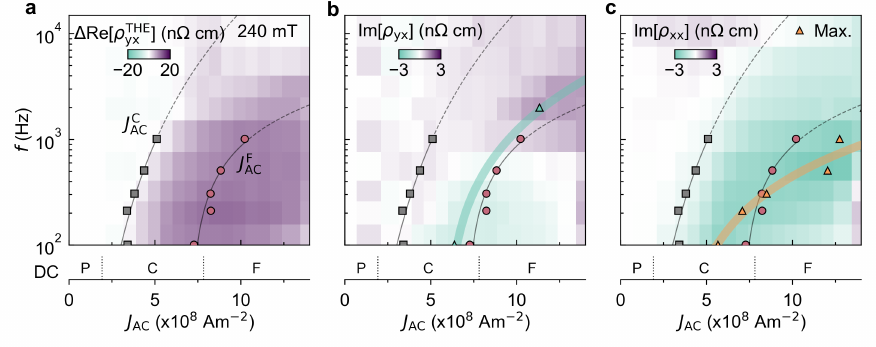}
    \caption{\textbf{Frequency and current dependence of emergent reactance:} \textbf{a}-\textbf{c} Colour maps of $\Delta$Re[$\rho_\mathrm{yx}^\mathrm{THE}$] (\textbf{a}), Im[$\rho_\mathrm{yx}$] (\textbf{b}), and Im[$\rho_\mathrm{xx}$] (\textbf{c}) on the $J_\mathrm{AC}$ - $f$  plane, measured at \SI{23}{\kelvin} for $B = $ \SI{0.24}{\tesla}. The creep threshold as characterized from  $\Delta$Re[$\rho_\mathrm{yx}^\mathrm{THE}$] is overlaid in each plot by the grey squares and red circles, respectively. The creep and flow thresholds in the DC case are indicated in the bottom of each panels. In (\textbf{b,c}) triangular markers denote the 	maximum and minimum of Im[$\rho_\mathrm{yx}$] and Im[$\rho_\mathrm{xx}$] in the $J_\mathrm{AC}$-$f$ plane. The teal and orange lines in (\textbf{b}) and (\textbf{c}) respectively show a guide of the maximum/minimum values for increasing frequency.}
    \label{f4}
\end{figure*}

Armed with the knowledge of the dynamical phases of the SkL in our device, we measured the AC current density dependence of the reactance signals to uncover their origins. For AC current measurements without $J_\mathrm{DC}$, the current passes through each sequential dynamical phases (Figs. \ref{f3}a and b). Consequently, the current density dependencies measured with DC and AC currents are not identical. However, since the voltage response is considerably influenced by the dynamical phase at the peak of the AC current, the dependence on the AC current amplitude provides valuable insights.
In Fig. \ref{f3}c, we present the change in the topological Hall resistivity $\Delta$Re[$\rho_\mathrm{yx}^\mathrm{THE}$] measured at a frequency of $f =$ \SI{101}{\hertz} as a function of $J_\mathrm{AC}$, demonstrating a similar profile to that of the DC bias current dependence (Fig. \ref{f2}b). From $\Delta$Re[$\rho_\mathrm{yx}^\mathrm{THE}$] in the SkL phase,  we determine the creep ($J_\mathrm{AC}^\mathrm{C}$) and flow ($J_\mathrm{AC}^\mathrm{F}$) thresholds for AC current (see Methods for the definition). At this low frequency, these values are close to $J_\mathrm{DC}^\mathrm{C}$ and $J_\mathrm{DC}^\mathrm{F}$. We note that a small difference between the threshold values for AC and DC current mainly arises from their frequency dependence (see the discussion below). In Figs. \ref{f3}d and e, we show the transverse and longitudinal reactance (Im[$\rho_\mathrm{yx}$] and Im[$\rho_\mathrm{xx}]$). While no reactance is observed for the helical and conical phases, nonzero reactance appears in the SkL phase. Notably, in the SkL phase, both the Hall and longitudinal reactance remain zero in the pinned region, reach their maximum values in the creep region, and then decrease in the flow region. In addition, in the case of DC current dependence, prominent reactance signals are also observed mainly in the creep region (see Supplementary Information). Therefore, these results suggest that the observed reactance is associated with the creep motion.

To provide further insight into the origin of these signals, we repeated such AC current dependent measurements at a range of frequencies, and plot colour maps of $\Delta$Re[$\rho_\mathrm{yx}^\mathrm{THE}$],  Im[$\rho_\mathrm{yx}]$ and Im[$\rho_\mathrm{xx}]$ in the frequency-current density plane in Figs. \ref{f4}(a-c). The creep and flow thresholds are indicated by solid lines. Extrapolating $J_\mathrm{AC}^\mathrm{C}$ and $J_\mathrm{AC}^\mathrm{F}$ to the low-frequency limit closely matches $J_\mathrm{DC}^\mathrm{C}$ and $J_\mathrm{DC}^\mathrm{F}$. In addition, with increasing frequency, $J_\mathrm{AC}^\mathrm{C}$ and $J_\mathrm{AC}^\mathrm{F}$ increase monotonically. This occurs due to freezing of the skyrmion motion at high frequencies, representative of low-frequency SkL dynamics. Both Im[$\rho_\mathrm{yx}$] and Im[$\rho_\mathrm{xx}$] (Figs. \ref{f4}b and c) appear above  $J_\mathrm{AC}^\mathrm{C}$, and reach their maximum mostly within the creep region (the triangular points in Figs. \ref{f4}b and c), implying that skyrmion dynamics in the creep region play an important role in both Im[$\rho_\mathrm{yx}$] and Im[$\rho_\mathrm{xx}$]. We note that the peak value of Im[$\rho_\mathrm{xx}$] shifts into the flow region at high frequencies. This originates from the AC current traversing the creep region before entering into the flow region. With increasing frequency, Im[$\rho_\mathrm{yx}$] changes sign from negative to positive around $f \approx$ \SI{500}{\hertz}, in stark contrast to Im[$\rho_\mathrm{xx}$], which maintains its sign (Fig. \ref{f4}c). This behaviour strongly suggests that, although they are both related to the SkL dynamics, they may have fundamentally different origins.  
 
Since Im[$\rho_\mathrm{yx}$] and Im[$\rho_\mathrm{xx}$] are observed in the moving SkL phase, it is natural to consider their origin in terms of emergent electric fields. The transverse reactance Im[$\rho_\mathrm{yx}$] can be explained by the inertial translational motion of the SkL. The velocity of the translational motion of the skyrmion $v_\mathrm{sk}$ obeys the Thiele equation:  $m_\mathrm{sk} \dot{\bm{v}}_{\rm{sk}}+\mathcal{G} \times({\bm{v}}_{\mathrm{e}}-\bm{v}_\mathrm{sk}) +  \mathcal{D} ({\beta\bm{v}}_{\rm{e}} -\alpha\bm{v}_\mathrm{sk}) + \bm{\nabla}V_\mathrm{pin} = 0$, where $\mathcal{G}$,  $\mathcal{D}$, $\beta$, $\alpha$, $V_\mathrm{pin}$, and $m_\mathrm{sk}$ are the gyro-coupling vector, dissipative force tensor, dimensionless constant characterizing the nonadiabatic electron spin dynamics, Gilbert damping constant, pinning potential, and the skyrmion mass, respectively. Here, the skyrmion mass arises because the deformation of skyrmions allows energy to be stored, which is renormalized into a mass-like term in the Thiele equation \cite{Thiele1973, Chutte2014}. Experimentally, inertial dynamics arising from this mass-like term have been reported in isolated skyrmion systems \cite{Buttner2015, Song2024} and in SkL \cite{Birch2024}. In the present case, since the skyrmion deformation occurs in the creep region, $m_\mathrm{sk}$ is nonzero in this region. Consequently, in the creep region, the time derivative of  $\bm{v}_\mathrm{Sk}$ and the nonlinear term $\bm{\nabla} V_\mathrm{pin}$ in the Thiele equation cause the phase of $\bm{v}_\mathrm{Sk}$ to shift relative to that of the input AC current. As a result, the phase of the emergent electric field induced by the translational motion of the SkL $\bm{e}_\mathrm{em} = -\bm{v}_\mathrm{Sk}\times\bm{b}_\mathrm{em}$ also shifts with respect to the input AC current, generating the transverse reactance component given by $\mathrm{Im}\rho_\mathrm{yx} = Pb_\mathrm{em}v''_\mathrm{Sk}/j_\mathrm{AC}$, where $\bm{v}''_\mathrm{sk}$ is the phase-shifted skyrmion velocity (see Supplementary Information). We note that $\bm{v}''_\mathrm{Sk}$ can be positive or negative, depending on the damping parameters, the shape of the potential, and the frequency, which is responsible for the sign change of Im[$\rho_\mathrm{yx}$] observed in the $B$-$T$ plane (Fig. \ref{f1}e) and the frequency-current plane (Fig. \ref{f4}b). In contrast, in the flow region, since $m_\mathrm{Sk}=0$ and $\bm{\nabla} V_\mathrm{pin}\approx 0 $  due to the absence of internal deformation and a relative reduction of the pinning force, the skyrmion velocity equals the electron velocity. Therefore, $\bm{e}_\mathrm{em} = -\bm{v}_\mathrm{Sk}\times\bm{b}_\mathrm{em}$ does not exhibit a phase shift, and thus the transverse reactance component vanishes.

In contrast, the longitudinal reactance cannot be explained by the inertial translational motion of the SkL. If the skyrmion velocity had a transverse component due to the skyrmion Hall effect, inertial translational SkL motion would generate a longitudinal reactance due to the projection of the emergent electric field along the longitudinal direction. This mechanism is proposed to explain the observed longitudinal reactance in \cite{Scheuchenpflug2025}. In this case, however, Im[$\rho_\mathrm{xx}$] and  Im[$\rho_\mathrm{yx}$] should have the same sign because the sign of the skyrmion Hall angle is determined by the topological charge and remains constant. Thus, this mechanism cannot explain why only Im[$\rho_\mathrm{yx}$] changes its sign in the $B$-$T$ plane (Figs. \ref{f1}e and g) and the frequency-current plane (Figs. \ref{f4}b and c). However, since the longitudinal reactance is also observed in the creep region, where the SkL moves while deforming, the dynamics and deformation of the SkL must play an important role. One plausible explanation is that the longitudinal reactance Im[$\rho_\mathrm{xx}$] arises directly from the internal deformation of the skyrmions themselves. The magnetic moment of the deformed SkL can be described as a superposition of three helices including its deformation as follows: 
\begin{align}
    \bm{m} &= m_z\bm{\hat{z}} + \sum_{i = a,b,c}\bm{m}_{i},\\
    \bm{m}_i &= m_h(\beta_i\bm{\hat{Q}}_i + \sqrt{1-\beta_i^2}\bm{l}_i),\\
    \bm{l}_i &= \bm{\hat{z}}\cos(\bm{Q}_i\cdot\bm{r}+\varphi_i) + (\bm{\hat{Q}}_i\times\bm{\hat{z}})\sin(\bm{Q}_i\cdot\bm{r} + \varphi_i),
\end{align}
where $\bm{Q}_i$ and $\bm{\hat{Q}}_i$ are the \textit{Q} vector for each helix and its unit vector. Here, $\varphi_i$ and $\beta_i$ are the phason and spin-tilting mode for each helix, respectively, which characterize the SkL deformation \cite{Tatara2014}. For a single-$Q$ helix, the longitudinal emergent reactance is induced by the excitation of these modes \cite{Nagaosa2019, Hoshino2018}. Likewise, their excitation in the SkL by its deformation in the creep regime are expected to generate a longitudinal emergent reactance. For example, when the spin-tilting mode parallel to the current direction ($\beta_{x}$) is excited, it can be shown that the longitudinal emergent electric field is proportional to its time derivative (${d\beta_{x}}/{dt}$), while the transverse reactance becomes zero, similar to the helix (see Supplementary Information).


Although in principle the helical and conical states can induce an emergent reactance, it was not observed in these phases. For the conical states, $\bm{Q}$ aligns parallel to $B$ in the present experimental setup, and explicitly perpendicular to $J$, forbidding excited phason and spin tilting modes \cite{Nagaosa2019}. For the helical state, $\bm{Q}$ prefers to align along the in-plane  \hkl<111> directions \cite{Ishikawa1976}, therefore it is possible to excite these modes and generate a longitudinal emergent reactance. However, since the threshold current density for the helical motion is much higher than that for the SkL \cite{Kimoto2025}, we expect no emergent electric field to be generated under the current densities investigated in the present work.
These reasons highlight the potential advantages of the SkL over other magnetic structures in generating the emergent reactance.

Our scenario clearly explains the temperature dependencies of Im[$\rho_\mathrm{yx}$] and Im[$\rho_\mathrm{xx}$](Figs. \ref{f1}e and g), both of which exhibit large signals in the centre of the SkL phase in the $B$-$T$ phase diagram. First, near the paramagnetic transition temperature (approximately \SI{25}-\SI{28}{\kelvin}), Im[$\rho_\mathrm{yx}$] at $J_\mathrm{AC} = $ \SI{6.81e8}{\ampere\per\meter\squared} falls below the noise level (Fig. \ref{f1}e). This occurs because the SkL remains pinned at this current density due to the increase in the critical current density required for creep motion at higher temperatures as reported in \cite{Schulz2012}. This is a natural extension of the Anderson-Kim theory \cite{Anderson1964}. Close to the fluctuation disordered boundary, strong thermal fluctuations soften the magnetic moment of the SkL, thereby enhancing the pinning effects. In fact, the topological Hall resistivity at \SI{26}{\kelvin} does not decrease even at  $J_\mathrm{AC} = $ \SI{6.81e8}{\ampere\per\meter\squared} (Supplementary Fig. S6), supporting the conclusion that the SkL remains pinned.  Accordingly, Im[$\rho_\mathrm{xx}$] is also substantially  reduced in this temperature range, while a small positive Im[$\rho_\mathrm{xx}$] $\approx$ \SI{0.25}{\nano\ohm\cm} is observed, which may result from an emergent reactance due to the current-induced deformation of the pinned SkL \cite{Furuta2023, Scheuchenpflug2025}, (Fig. \ref{f1}g) (see Supplementary Information). Secondly, Im[$\rho_\mathrm{yx}$] and Im[$\rho_\mathrm{xx}$] also become small at low temperatures below \SI{22}{\kelvin}. In this temperature range, the topological Hall resistivity measured at low current density is smaller than that in the centre of the SkL phase, indicating that the skyrmion state is metastable and coexists with the conical phase, which leads to small Im[$\rho_\mathrm{yx}$] and Im[$\rho_\mathrm{xx}$] (see Supplementary Information). 

In summary, we have demonstrated both transverse and longitudinal emergent reactance in the creep motion regime of the SkL. For both cases, the deformation of the SkL plays a crucial role, indirectly for transverse reactance and directly for longitudinal reactance. We attribute the transverse reactance to the phase shift of the emergent electric field resulting from inertial translational motion of the SkL, which arises from the renormalisation of its deformation into an effective skyrmion mass. In contrast, the longitudinal emergent reactance is attributed to the emergent electric field induced by the excitation of the phason and spin tilting modes during the deformation of the SkL. Our work highlights the importance of internal deformation degrees of freedom in skyrmions for the generation of emergent electric field, an aspect that has been overlooked. Notably, the low critical current densities required for SkL creep motion may provide an advantage over helical and domain wall systems in generating emergent reactance \cite{Kitaori2021, Kitaori2024}. 
	

\vfill
\newpage

\section*{Methods}
\subsubsection{Sample preparation and device fabrication}
Single crystals of \ce{MnSi} were grown by the Czochralski method and oriented using a Laue x-ray camera. A FEI Helios 5UX focused ion beam instrument was used to fabricate the thin-plate devices from the single crystal. First, a large slab of material was created by means of \ce{Ga} ion milling. Then, the slab was lifted from the single crystal and attached to a \ce{Cu} TEM grid using an EasyLift micromanipulator. Subsequently, the slab was thinned to below \SI{1}{\micro\meter} and shaped using \ce{Ga} ion milling processes. Au electrical contacts were deposited onto a \ce{CaF2} substrate using maskless UV lithography and electron beam evaporation. The shaped \ce{MnSi} thin plate was then mounted onto the \ce{CaF2} substrate using the micromanipulator and \ce{Pt} deposition so that the sample itself directly touched the \ce{Au} electrical contacts. Finally, the sample was shaped into the Hall bar geometry via \ce{Ga} ion milling. In all measurements, the direction of the current $J$ is oriented along the long edge of the device, while the externally applied magnetic field $B$ is applied out-of-plane. Both $J$ and $B$ are oriented along mutually perpendicular \hkl<100> type crystal directions. 

\subsubsection{Transport measurements}
All transport measurements were performed within a Quantum Design Physical Properties Measurement System (PPMS). The \ce{MnSi} devices were mounted on a standard PPMS puck using GE varnish. \ce{Cu} wires were used to contact the substrate with a combination of solder and colloidal silver paste. Measurements of the longitudinal and Hall impedances were taken simultaneously using the lock-in technique (Stanford Instruments SR830) and were symmetrized and anti-symmetrized with respect to the magnetic field in the conventional manner i.e. $\rho_\mathrm{yx}^\mathrm{asym} = [\rho_\mathrm{yx}(+B) - \rho_\mathrm{yx}(-B)]/2$ and $\rho_\mathrm{xx}^\mathrm{sym} = [\rho_\mathrm{xx}(+B) + \rho_\mathrm{xx}(-B)]/2$. For the AC+DC measurements, a small AC oscillation was superimposed on a large DC bias using a Keithley 6221 current source. The AC peak amplitude was set to 1.74$\times$10$^8$\SI{}{\ampere\per\meter\squared} at a frequency of \SI{509}{\hertz}, and the output TTL pulse was used as the lock-in reference signal. In this case we technically measure the gradient of the I-V curve according to $\frac{dV}{dI}$, which we convert into the differential resistivity following $\rho^\mathrm{diff} = t\frac{dV}{dI}$, where $t$ denotes the sample thickness.

\subsubsection{Determination of magnetic phase boundaries}\label{PhaseBoundaries}
The phase boundaries are calculated for measurements taken with a low current density of $J_\mathrm{AC} = $ \SI{4.26e8}{\ampere\per\meter\squared} such as to capture the equilibrium skyrmion phase, avoiding potential expansions due to current induced skyrmion generation and Joule heating effects. The higher current density facilitates the extraction of the phase boundaries while being qualitatively similar to those observed at lower $J_\mathrm{AC}$ (See supplementary information). Considering each isothermal magnetic field sweep individually, we first subtract the ordinary Hall effect by fitting a straight line to the resistivity where $|B| > $ \SI{0.7}{\tesla}. After subtraction, each magnetic phase (helical, conical, SkL, field-polarised) displays a different gradient in Re[$\rho_\mathrm{yx} - \rho_\mathrm{yx}^\mathrm{OHE}$] versus $B$. Accordingly, to estimate the phase boundaries, we first fit a straight line to the approximate centre of each phase, determined by eye. We then define the estimated phase boundary as the intersection between adjacent extrapolated fits. The solid lines between helical to conical $B_\mathrm{H}$ (green triangles in Fig. \ref{f1}(c)), and conical to field polarised $B_\mathrm{c}$ (grey squares in Fig. \ref{f1}(c)) boundaries are fit according to the following phenomenological order parameter equation:
\begin{equation}
	B_\mathrm{H,c}= \begin{cases} 
		0 & T \geq T_\mathrm{C} \\
		A(1-\frac{T}{T_\mathrm{C}})^\beta & T \leq T_\mathrm{C}
	\end{cases},
\end{equation}
where $A, \beta$ are fitting parameters, and $T_\mathrm{C}$ is a fitting parameter which refers to the transition temperature. For the H transition, we determine $T_\mathrm{C} = $ \SI{28.01(2)}{\kelvin}. The solid line for the conical to field polarised transition is determined by fitting the grey squares and fixing $T_\mathrm{C} = $ \SI{28.01(2)}{\kelvin}. Allowing $T_\mathrm{C}$ to vary during this transition leads to $T_\mathrm{C} = $ \SI{29.9(5)}{\kelvin}. The difference in these two transition temperatures accurately captures the fluctuation disordered regime, closely replicating the phase boundaries in bulk \ce{MnSi} \cite{Bauer2012}. 

\subsubsection{Determination of creep and flow thresholds}
To determine the creep ($J^C$) and flow ($J^F$) thresholds, we first fit $\Delta$Re[$\rho_\mathrm{yx}$] according to the equation
\begin{equation}
    \Delta\mathrm{Re}\left[\rho_\mathrm{yx}^\mathrm{THE}\right] = a\left[\frac{a}{\sqrt{\pi}}\int_0^{\frac{J-b}{c}}e^{-t^2}dt+1\right],
    \label{Erf}
\end{equation}
where $a$,$b$, and $c$ are fitting parameters. This equation provides a good phenomenological approximation of the profile of the current density dependence of $\Delta$Re[$\rho_\mathrm{yx}^\mathrm{THE}$]. To obtain the creep threshold, we fit a straight line to the pinned regime and creep transition and determine their intersect. The flow regime is defined as the point at which $\Delta$Re[$\rho_\mathrm{yx}^\mathrm{THE}$] saturates. An example of this procedure is shown in Supplementary Fig. S7. \\

\subsubsection{Calculation of electron velocity}
The electron velocity in Fig. \ref{f2}(c) was calculated from the ordinary Hall coefficient $R_0$ according to the classical Drude model such that
\begin{equation}
	v_e = R_HJ_\mathrm{DC} = \frac{J_\mathrm{DC}}{ne},
\end{equation}
where $n$ is the electron number density and $e$ is the electron charge. $R_H$ was obtained by fitting a straight line to the resistivity where $|B| >$ \SI{0.7}{\tesla}. An example is shown in Supplementary Fig. S2.

\section*{Acknowledgements}
We thank the CEMS Semiconductor Science Research Support Team for the assistance in the use of cleanroom facilities. We thank X. Z. Yu and her team for the use of the FEI Helios 5UX focused ion beam system.  This work was supported by JSPS KAKENHI Grant Numbers 23H05431, 24K00566, 24H00197, and 24H02231 and JST PRESTO, Grant Number JPMJPR235B. M. T. L was supported by the JSPS Summer Program 2024 in collaboration with SOKENDAI, who funded M.T.L's placement at RIKEN, Japan. N.N. was supported by the RIKEN TRIP initiative.  

\section*{Contributions}
M. T. L, M. T. B, D. A. V, P. D. H, and T. Y conceived the project. A. K. and Y. Ta. grew the bulk \ce{MnSi} single crystal. M. T. L fabricated the focused ion beam devices with support from M.T.B. M.T.L performed the measurements and analysis with support from M. T. B and T. Y; T. Y and N. N developed the phenomenological models. M. T. L, M. T. B, N. N, Y. To, and T. Y interpreted the data and wrote the paper, along with contributions from all other co-authors.

\vfill

\bibliography{bib}

\begin{thebibliography}{13}%
\makeatletter
\providecommand \@ifxundefined [1]{%
 \@ifx{#1\undefined}
}%
\providecommand \@ifnum [1]{%
 \ifnum #1\expandafter \@firstoftwo
 \else \expandafter \@secondoftwo
 \fi
}%
\providecommand \@ifx [1]{%
 \ifx #1\expandafter \@firstoftwo
 \else \expandafter \@secondoftwo
 \fi
}%
\providecommand \natexlab [1]{#1}%
\providecommand \enquote  [1]{``#1''}%
\providecommand \bibnamefont  [1]{#1}%
\providecommand \bibfnamefont [1]{#1}%
\providecommand \citenamefont [1]{#1}%
\providecommand \href@noop [0]{\@secondoftwo}%
\providecommand \href [0]{\begingroup \@sanitize@url \@href}%
\providecommand \@href[1]{\@@startlink{#1}\@@href}%
\providecommand \@@href[1]{\endgroup#1\@@endlink}%
\providecommand \@sanitize@url [0]{\catcode `\\12\catcode `\$12\catcode
  `\&12\catcode `\#12\catcode `\^12\catcode `\_12\catcode `\%12\relax}%
\providecommand \@@startlink[1]{}%
\providecommand \@@endlink[0]{}%
\providecommand \url  [0]{\begingroup\@sanitize@url \@url }%
\providecommand \@url [1]{\endgroup\@href {#1}{\urlprefix }}%
\providecommand \urlprefix  [0]{URL }%
\providecommand \Eprint [0]{\href }%
\providecommand \doibase [0]{https://doi.org/}%
\providecommand \selectlanguage [0]{\@gobble}%
\providecommand \bibinfo  [0]{\@secondoftwo}%
\providecommand \bibfield  [0]{\@secondoftwo}%
\providecommand \translation [1]{[#1]}%
\providecommand \BibitemOpen [0]{}%
\providecommand \bibitemStop [0]{}%
\providecommand \bibitemNoStop [0]{.\EOS\space}%
\providecommand \EOS [0]{\spacefactor3000\relax}%
\providecommand \BibitemShut  [1]{\csname bibitem#1\endcsname}%
\let\auto@bib@innerbib\@empty
\bibitem [{\citenamefont {Choi}\ \emph {et~al.}(2024)\citenamefont {Choi},
  \citenamefont {Lee}, \citenamefont {Yang}, \citenamefont {Kang},
  \citenamefont {Park}, \citenamefont {Park},\ and\ \citenamefont
  {Kim}}]{Choi2024}%
  \BibitemOpen
  \bibfield  {author} {\bibinfo {author} {\bibfnamefont {J.}~\bibnamefont
  {Choi}}, \bibinfo {author} {\bibfnamefont {G.-H.}\ \bibnamefont {Lee}},
  \bibinfo {author} {\bibfnamefont {J.}~\bibnamefont {Yang}}, \bibinfo {author}
  {\bibfnamefont {J.}~\bibnamefont {Kang}}, \bibinfo {author} {\bibfnamefont
  {B.-G.}\ \bibnamefont {Park}}, \bibinfo {author} {\bibfnamefont {A.~M.~G.}\
  \bibnamefont {Park}},\ and\ \bibinfo {author} {\bibfnamefont {K.-J.}\
  \bibnamefont {Kim}},\ }\bibfield  {title} {\bibinfo {title} {Questioning the
  validity of spintronic inductors: Potential artifacts in emergent
  inductance},\ }\href {https://doi.org/10.1063/5.0227569} {\bibfield
  {journal} {\bibinfo  {journal} {Applied Physics Letters}\ }\textbf {\bibinfo
  {volume} {125}},\ \bibinfo {pages} {192403} (\bibinfo {year}
  {2024})}\BibitemShut {NoStop}%
\bibitem [{\citenamefont {Bauer}\ and\ \citenamefont
  {Pfleiderer}(2012)}]{Bauer2012}%
  \BibitemOpen
  \bibfield  {author} {\bibinfo {author} {\bibfnamefont {A.}~\bibnamefont
  {Bauer}}\ and\ \bibinfo {author} {\bibfnamefont {C.}~\bibnamefont
  {Pfleiderer}},\ }\bibfield  {title} {\bibinfo {title} {Magnetic phase diagram
  of {MnSi} inferred from magnetization and {AC} susceptibility},\ }\href
  {https://doi.org/10.1103/PhysRevB.85.214418} {\bibfield  {journal} {\bibinfo
  {journal} {Physical Review B}\ }\textbf {\bibinfo {volume} {85}},\ \bibinfo
  {pages} {214418} (\bibinfo {year} {2012})}\BibitemShut {NoStop}%
\bibitem [{\citenamefont {Birch}\ \emph {et~al.}(2024)\citenamefont {Birch},
  \citenamefont {Belopolski}, \citenamefont {Fujishiro}, \citenamefont
  {Kawamura}, \citenamefont {Kikkawa}, \citenamefont {Taguchi}, \citenamefont
  {Hirschberger}, \citenamefont {Nagaosa},\ and\ \citenamefont
  {Tokura}}]{Birch2024}%
  \BibitemOpen
  \bibfield  {author} {\bibinfo {author} {\bibfnamefont {M.~T.}\ \bibnamefont
  {Birch}}, \bibinfo {author} {\bibfnamefont {I.}~\bibnamefont {Belopolski}},
  \bibinfo {author} {\bibfnamefont {Y.}~\bibnamefont {Fujishiro}}, \bibinfo
  {author} {\bibfnamefont {M.}~\bibnamefont {Kawamura}}, \bibinfo {author}
  {\bibfnamefont {A.}~\bibnamefont {Kikkawa}}, \bibinfo {author} {\bibfnamefont
  {Y.}~\bibnamefont {Taguchi}}, \bibinfo {author} {\bibfnamefont
  {M.}~\bibnamefont {Hirschberger}}, \bibinfo {author} {\bibfnamefont
  {N.}~\bibnamefont {Nagaosa}},\ and\ \bibinfo {author} {\bibfnamefont
  {Y.}~\bibnamefont {Tokura}},\ }\bibfield  {title} {\bibinfo {title} {Dynamic
  transition and galilean relativity of current-driven skyrmions},\ }\href
  {https://doi.org/10.1038/s41586-024-07859-2} {\bibfield  {journal} {\bibinfo
  {journal} {Nature}\ }\textbf {\bibinfo {volume} {633}},\ \bibinfo {pages}
  {554} (\bibinfo {year} {2024})}\BibitemShut {NoStop}%
\bibitem [{\citenamefont {Schulz}\ \emph {et~al.}(2012)\citenamefont {Schulz},
  \citenamefont {Ritz}, \citenamefont {Bauer}, \citenamefont {Halder},
  \citenamefont {Wagner}, \citenamefont {Franz}, \citenamefont {Pfleiderer},
  \citenamefont {Everschor}, \citenamefont {Garst},\ and\ \citenamefont
  {Rosch}}]{Schulz2012}%
  \BibitemOpen
  \bibfield  {author} {\bibinfo {author} {\bibfnamefont {T.}~\bibnamefont
  {Schulz}}, \bibinfo {author} {\bibfnamefont {R.}~\bibnamefont {Ritz}},
  \bibinfo {author} {\bibfnamefont {A.}~\bibnamefont {Bauer}}, \bibinfo
  {author} {\bibfnamefont {M.}~\bibnamefont {Halder}}, \bibinfo {author}
  {\bibfnamefont {M.}~\bibnamefont {Wagner}}, \bibinfo {author} {\bibfnamefont
  {C.}~\bibnamefont {Franz}}, \bibinfo {author} {\bibfnamefont
  {C.}~\bibnamefont {Pfleiderer}}, \bibinfo {author} {\bibfnamefont
  {K.}~\bibnamefont {Everschor}}, \bibinfo {author} {\bibfnamefont
  {M.}~\bibnamefont {Garst}},\ and\ \bibinfo {author} {\bibfnamefont
  {A.}~\bibnamefont {Rosch}},\ }\bibfield  {title} {\bibinfo {title} {Emergent
  electrodynamics of skyrmions in a chiral magnet},\ }\href
  {https://doi.org/10.1038/nphys2231} {\bibfield  {journal} {\bibinfo
  {journal} {Nature Physics}\ }\textbf {\bibinfo {volume} {8}},\ \bibinfo
  {pages} {301} (\bibinfo {year} {2012})}\BibitemShut {NoStop}%
\bibitem [{\citenamefont {Iwasaki}\ \emph
  {et~al.}(2013{\natexlab{a}})\citenamefont {Iwasaki}, \citenamefont
  {Mochizuki},\ and\ \citenamefont {Nagaosa}}]{Iwasaki2013}%
  \BibitemOpen
  \bibfield  {author} {\bibinfo {author} {\bibfnamefont {J.}~\bibnamefont
  {Iwasaki}}, \bibinfo {author} {\bibfnamefont {M.}~\bibnamefont {Mochizuki}},\
  and\ \bibinfo {author} {\bibfnamefont {N.}~\bibnamefont {Nagaosa}},\
  }\bibfield  {title} {\bibinfo {title} {Universal current-velocity relation of
  skyrmion motion in chiral magnets},\ }\href
  {https://doi.org/10.1038/ncomms2442} {\bibfield  {journal} {\bibinfo
  {journal} {Nature Communications}\ }\textbf {\bibinfo {volume} {4}},\
  \bibinfo {pages} {1463} (\bibinfo {year} {2013}{\natexlab{a}})}\BibitemShut
  {NoStop}%
\bibitem [{\citenamefont {Iwasaki}\ \emph
  {et~al.}(2013{\natexlab{b}})\citenamefont {Iwasaki}, \citenamefont
  {Mochizuki},\ and\ \citenamefont {Nagaosa}}]{Iwasaki2013_2}%
  \BibitemOpen
  \bibfield  {author} {\bibinfo {author} {\bibfnamefont {J.}~\bibnamefont
  {Iwasaki}}, \bibinfo {author} {\bibfnamefont {M.}~\bibnamefont {Mochizuki}},\
  and\ \bibinfo {author} {\bibfnamefont {N.}~\bibnamefont {Nagaosa}},\
  }\bibfield  {title} {\bibinfo {title} {Current-induced skyrmion dynamics in
  constricted geometries},\ }\href {https://doi.org/10.1038/nnano.2013.176}
  {\bibfield  {journal} {\bibinfo  {journal} {Nature Nanotechnology}\ }\textbf
  {\bibinfo {volume} {8}},\ \bibinfo {pages} {742} (\bibinfo {year}
  {2013}{\natexlab{b}})}\BibitemShut {NoStop}%
\bibitem [{\citenamefont {Tatara}\ and\ \citenamefont
  {Fukuyama}(2014)}]{Tatara2014}%
  \BibitemOpen
  \bibfield  {author} {\bibinfo {author} {\bibfnamefont {G.}~\bibnamefont
  {Tatara}}\ and\ \bibinfo {author} {\bibfnamefont {H.}~\bibnamefont
  {Fukuyama}},\ }\bibfield  {title} {\bibinfo {title} {Phasons and excitations
  in skyrmion lattice},\ }\href {https://doi.org/10.7566/JPSJ.83.104711}
  {\bibfield  {journal} {\bibinfo  {journal} {Journal of the Physical Society
  of Japan}\ }\textbf {\bibinfo {volume} {83}},\ \bibinfo {pages} {104711}
  (\bibinfo {year} {2014})}\BibitemShut {NoStop}%
\bibitem [{\citenamefont {Nagaosa}(2019)}]{Nagaosa2019}%
  \BibitemOpen
  \bibfield  {author} {\bibinfo {author} {\bibfnamefont {N.}~\bibnamefont
  {Nagaosa}},\ }\bibfield  {title} {\bibinfo {title} {Emergent inductor by
  spiral magnets},\ }\href {https://doi.org/10.7567/1347-4065/ab5294}
  {\bibfield  {journal} {\bibinfo  {journal} {Japanese Journal of Applied
  Physics}\ }\textbf {\bibinfo {volume} {58}},\ \bibinfo {pages} {120909}
  (\bibinfo {year} {2019})}\BibitemShut {NoStop}%
\bibitem [{\citenamefont {Kurebayashi}\ and\ \citenamefont
  {Nagaosa}(2021)}]{Kurebayashi2021}%
  \BibitemOpen
  \bibfield  {author} {\bibinfo {author} {\bibfnamefont {D.}~\bibnamefont
  {Kurebayashi}}\ and\ \bibinfo {author} {\bibfnamefont {N.}~\bibnamefont
  {Nagaosa}},\ }\bibfield  {title} {\bibinfo {title} {Electromagnetic response
  in spiral magnets and emergent inductance},\ }\href
  {https://doi.org/10.1038/s42005-021-00765-3} {\bibfield  {journal} {\bibinfo
  {journal} {Communications Physics}\ }\textbf {\bibinfo {volume} {4}},\
  \bibinfo {pages} {260} (\bibinfo {year} {2021})}\BibitemShut {NoStop}%
\bibitem [{\citenamefont {Lemesh}\ \emph {et~al.}(2018)\citenamefont {Lemesh},
  \citenamefont {Litzius}, \citenamefont {B{\"o}ttcher}, \citenamefont
  {Bassirian}, \citenamefont {Kerber}, \citenamefont {Heinze}, \citenamefont
  {Z{\'a}zvorka}, \citenamefont {B{\"u}ttner}, \citenamefont {Caretta},
  \citenamefont {Mann}, \citenamefont {Weigand}, \citenamefont {Finizio},
  \citenamefont {Raabe}, \citenamefont {Im}, \citenamefont {Stoll},
  \citenamefont {Sch{\"u}tz}, \citenamefont {Dup{\'e}}, \citenamefont
  {Kl{\"a}ui},\ and\ \citenamefont {Beach}}]{Lemesh2018}%
  \BibitemOpen
  \bibfield  {author} {\bibinfo {author} {\bibfnamefont {I.}~\bibnamefont
  {Lemesh}}, \bibinfo {author} {\bibfnamefont {K.}~\bibnamefont {Litzius}},
  \bibinfo {author} {\bibfnamefont {M.}~\bibnamefont {B{\"o}ttcher}}, \bibinfo
  {author} {\bibfnamefont {P.}~\bibnamefont {Bassirian}}, \bibinfo {author}
  {\bibfnamefont {N.}~\bibnamefont {Kerber}}, \bibinfo {author} {\bibfnamefont
  {D.}~\bibnamefont {Heinze}}, \bibinfo {author} {\bibfnamefont
  {J.}~\bibnamefont {Z{\'a}zvorka}}, \bibinfo {author} {\bibfnamefont
  {F.}~\bibnamefont {B{\"u}ttner}}, \bibinfo {author} {\bibfnamefont
  {L.}~\bibnamefont {Caretta}}, \bibinfo {author} {\bibfnamefont
  {M.}~\bibnamefont {Mann}}, \bibinfo {author} {\bibfnamefont {M.}~\bibnamefont
  {Weigand}}, \bibinfo {author} {\bibfnamefont {S.}~\bibnamefont {Finizio}},
  \bibinfo {author} {\bibfnamefont {J.}~\bibnamefont {Raabe}}, \bibinfo
  {author} {\bibfnamefont {M.-Y.}\ \bibnamefont {Im}}, \bibinfo {author}
  {\bibfnamefont {H.}~\bibnamefont {Stoll}}, \bibinfo {author} {\bibfnamefont
  {G.}~\bibnamefont {Sch{\"u}tz}}, \bibinfo {author} {\bibfnamefont
  {B.}~\bibnamefont {Dup{\'e}}}, \bibinfo {author} {\bibfnamefont
  {M.}~\bibnamefont {Kl{\"a}ui}},\ and\ \bibinfo {author} {\bibfnamefont
  {G.~S.~D.}\ \bibnamefont {Beach}},\ }\bibfield  {title} {\bibinfo {title}
  {Current-induced skyrmion generation through morphological thermal
  transitions in chiral ferromagnetic heterostructures},\ }\href
  {https://doi.org/https://doi.org/10.1002/adma.201805461} {\bibfield
  {journal} {\bibinfo  {journal} {Advanced Materials}\ }\textbf {\bibinfo
  {volume} {30}},\ \bibinfo {pages} {1805461} (\bibinfo {year}
  {2018})}\BibitemShut {NoStop}%
\bibitem [{\citenamefont {Furuta}\ \emph {et~al.}(2023)\citenamefont {Furuta},
  \citenamefont {Koshibae},\ and\ \citenamefont {Kagawa}}]{Furuta2023}%
  \BibitemOpen
  \bibfield  {author} {\bibinfo {author} {\bibfnamefont {S.}~\bibnamefont
  {Furuta}}, \bibinfo {author} {\bibfnamefont {W.}~\bibnamefont {Koshibae}},\
  and\ \bibinfo {author} {\bibfnamefont {F.}~\bibnamefont {Kagawa}},\
  }\bibfield  {title} {\bibinfo {title} {Symmetry of the emergent inductance
  tensor exhibited by magnetic textures},\ }\href
  {https://doi.org/10.1038/s44306-023-00001-4} {\bibfield  {journal} {\bibinfo
  {journal} {npj Spintronics}\ }\textbf {\bibinfo {volume} {1}},\ \bibinfo
  {pages} {3} (\bibinfo {year} {2023})}\BibitemShut {NoStop}%
\bibitem [{\citenamefont {Furuta}\ \emph {et~al.}(2024)\citenamefont {Furuta},
  \citenamefont {Koshibae}, \citenamefont {Matsuura}, \citenamefont {Abe},
  \citenamefont {Wang}, \citenamefont {Zhou}, \citenamefont {Arima},\ and\
  \citenamefont {Kagawa}}]{Furata2024}%
  \BibitemOpen
  \bibfield  {author} {\bibinfo {author} {\bibfnamefont {S.}~\bibnamefont
  {Furuta}}, \bibinfo {author} {\bibfnamefont {W.}~\bibnamefont {Koshibae}},
  \bibinfo {author} {\bibfnamefont {K.}~\bibnamefont {Matsuura}}, \bibinfo
  {author} {\bibfnamefont {N.}~\bibnamefont {Abe}}, \bibinfo {author}
  {\bibfnamefont {F.}~\bibnamefont {Wang}}, \bibinfo {author} {\bibfnamefont
  {S.}~\bibnamefont {Zhou}}, \bibinfo {author} {\bibfnamefont {T.-h.}\
  \bibnamefont {Arima}},\ and\ \bibinfo {author} {\bibfnamefont
  {F.}~\bibnamefont {Kagawa}},\ }\bibfield  {title} {\bibinfo {title}
  {Reconsidering nonlinear emergent inductance: Time-varying joule heating and
  its impact on {AC} electrical response},\ }\href
  {https://doi.org/10.1103/PhysRevB.110.174402} {\bibfield  {journal} {\bibinfo
   {journal} {Physical Review B}\ }\textbf {\bibinfo {volume} {110}},\ \bibinfo
  {pages} {174402} (\bibinfo {year} {2024})}\BibitemShut {NoStop}%
\bibitem [{\citenamefont {Sato}\ \emph {et~al.}(2022)\citenamefont {Sato},
  \citenamefont {Koshibae}, \citenamefont {Kikkawa}, \citenamefont {Taguchi},
  \citenamefont {Nagaosa}, \citenamefont {Tokura},\ and\ \citenamefont
  {Kagawa}}]{Sato2022}%
  \BibitemOpen
  \bibfield  {author} {\bibinfo {author} {\bibfnamefont {T.}~\bibnamefont
  {Sato}}, \bibinfo {author} {\bibfnamefont {W.}~\bibnamefont {Koshibae}},
  \bibinfo {author} {\bibfnamefont {A.}~\bibnamefont {Kikkawa}}, \bibinfo
  {author} {\bibfnamefont {Y.}~\bibnamefont {Taguchi}}, \bibinfo {author}
  {\bibfnamefont {N.}~\bibnamefont {Nagaosa}}, \bibinfo {author} {\bibfnamefont
  {Y.}~\bibnamefont {Tokura}},\ and\ \bibinfo {author} {\bibfnamefont
  {F.}~\bibnamefont {Kagawa}},\ }\bibfield  {title} {\bibinfo {title}
  {Nonthermal current-induced transition from skyrmion lattice to
  nontopological magnetic phase in spatially confined {MnSi}},\ }\href
  {https://doi.org/10.1103/PhysRevB.106.144425} {\bibfield  {journal} {\bibinfo
   {journal} {Physical Review B}\ }\textbf {\bibinfo {volume} {106}},\ \bibinfo
  {pages} {144425} (\bibinfo {year} {2022})}\BibitemShut {NoStop}%
\end{thebibliography}%


\begin{thebibliography}{46}%
\makeatletter
\providecommand \@ifxundefined [1]{%
 \@ifx{#1\undefined}
}%
\providecommand \@ifnum [1]{%
 \ifnum #1\expandafter \@firstoftwo
 \else \expandafter \@secondoftwo
 \fi
}%
\providecommand \@ifx [1]{%
 \ifx #1\expandafter \@firstoftwo
 \else \expandafter \@secondoftwo
 \fi
}%
\providecommand \natexlab [1]{#1}%
\providecommand \enquote  [1]{``#1''}%
\providecommand \bibnamefont  [1]{#1}%
\providecommand \bibfnamefont [1]{#1}%
\providecommand \citenamefont [1]{#1}%
\providecommand \href@noop [0]{\@secondoftwo}%
\providecommand \href [0]{\begingroup \@sanitize@url \@href}%
\providecommand \@href[1]{\@@startlink{#1}\@@href}%
\providecommand \@@href[1]{\endgroup#1\@@endlink}%
\providecommand \@sanitize@url [0]{\catcode `\\12\catcode `\$12\catcode
  `\&12\catcode `\#12\catcode `\^12\catcode `\_12\catcode `\%12\relax}%
\providecommand \@@startlink[1]{}%
\providecommand \@@endlink[0]{}%
\providecommand \url  [0]{\begingroup\@sanitize@url \@url }%
\providecommand \@url [1]{\endgroup\@href {#1}{\urlprefix }}%
\providecommand \urlprefix  [0]{URL }%
\providecommand \Eprint [0]{\href }%
\providecommand \doibase [0]{https://doi.org/}%
\providecommand \selectlanguage [0]{\@gobble}%
\providecommand \bibinfo  [0]{\@secondoftwo}%
\providecommand \bibfield  [0]{\@secondoftwo}%
\providecommand \translation [1]{[#1]}%
\providecommand \BibitemOpen [0]{}%
\providecommand \bibitemStop [0]{}%
\providecommand \bibitemNoStop [0]{.\EOS\space}%
\providecommand \EOS [0]{\spacefactor3000\relax}%
\providecommand \BibitemShut  [1]{\csname bibitem#1\endcsname}%
\let\auto@bib@innerbib\@empty
\bibitem [{\citenamefont {Berry}(1984)}]{Berry1984}%
  \BibitemOpen
  \bibfield  {author} {\bibinfo {author} {\bibfnamefont {M.~V.}\ \bibnamefont
  {Berry}},\ }\bibfield  {title} {\bibinfo {title} {Quantal phase factors
  accompanying adiabatic changes},\ }\href
  {https://api.semanticscholar.org/CorpusID:46623507} {\bibfield  {journal}
  {\bibinfo  {journal} {Proceedings of the Royal Society of London. A.
  Mathematical and Physical Sciences}\ }\textbf {\bibinfo {volume} {392}},\
  \bibinfo {pages} {45 } (\bibinfo {year} {1984})}\BibitemShut {NoStop}%
\bibitem [{\citenamefont {Nagaosa}\ \emph {et~al.}(2010)\citenamefont
  {Nagaosa}, \citenamefont {Sinova}, \citenamefont {Onoda}, \citenamefont
  {MacDonald},\ and\ \citenamefont {Ong}}]{Nagaosa2010}%
  \BibitemOpen
  \bibfield  {author} {\bibinfo {author} {\bibfnamefont {N.}~\bibnamefont
  {Nagaosa}}, \bibinfo {author} {\bibfnamefont {J.}~\bibnamefont {Sinova}},
  \bibinfo {author} {\bibfnamefont {S.}~\bibnamefont {Onoda}}, \bibinfo
  {author} {\bibfnamefont {A.~H.}\ \bibnamefont {MacDonald}},\ and\ \bibinfo
  {author} {\bibfnamefont {N.~P.}\ \bibnamefont {Ong}},\ }\bibfield  {title}
  {\bibinfo {title} {Anomalous {H}all effect},\ }\href
  {https://doi.org/10.1103/RevModPhys.82.1539} {\bibfield  {journal} {\bibinfo
  {journal} {Reviews of Modern Physics}\ }\textbf {\bibinfo {volume} {82}},\
  \bibinfo {pages} {1539} (\bibinfo {year} {2010})}\BibitemShut {NoStop}%
\bibitem [{\citenamefont {Zhang}\ \emph {et~al.}(2005)\citenamefont {Zhang},
  \citenamefont {Tan}, \citenamefont {Stormer},\ and\ \citenamefont
  {Kim}}]{Zhang2005}%
  \BibitemOpen
  \bibfield  {author} {\bibinfo {author} {\bibfnamefont {Y.}~\bibnamefont
  {Zhang}}, \bibinfo {author} {\bibfnamefont {Y.-W.}\ \bibnamefont {Tan}},
  \bibinfo {author} {\bibfnamefont {H.~L.}\ \bibnamefont {Stormer}},\ and\
  \bibinfo {author} {\bibfnamefont {P.}~\bibnamefont {Kim}},\ }\bibfield
  {title} {\bibinfo {title} {Experimental observation of the quantum {H}all
  effect and {B}erry's phase in graphene},\ }\href
  {https://doi.org/10.1038/nature04235} {\bibfield  {journal} {\bibinfo
  {journal} {Nature}\ }\textbf {\bibinfo {volume} {438}},\ \bibinfo {pages}
  {201} (\bibinfo {year} {2005})}\BibitemShut {NoStop}%
\bibitem [{\citenamefont {Qi}\ and\ \citenamefont {Zhang}(2011)}]{Qi2011}%
  \BibitemOpen
  \bibfield  {author} {\bibinfo {author} {\bibfnamefont {X.-L.}\ \bibnamefont
  {Qi}}\ and\ \bibinfo {author} {\bibfnamefont {S.-C.}\ \bibnamefont {Zhang}},\
  }\bibfield  {title} {\bibinfo {title} {Topological insulators and
  superconductors},\ }\href {https://doi.org/10.1103/RevModPhys.83.1057}
  {\bibfield  {journal} {\bibinfo  {journal} {Reviews of Modern Physics}\
  }\textbf {\bibinfo {volume} {83}},\ \bibinfo {pages} {1057} (\bibinfo {year}
  {2011})}\BibitemShut {NoStop}%
\bibitem [{\citenamefont {Armitage}\ \emph {et~al.}(2018)\citenamefont
  {Armitage}, \citenamefont {Mele},\ and\ \citenamefont
  {Vishwanath}}]{Armitage2018}%
  \BibitemOpen
  \bibfield  {author} {\bibinfo {author} {\bibfnamefont {N.~P.}\ \bibnamefont
  {Armitage}}, \bibinfo {author} {\bibfnamefont {E.~J.}\ \bibnamefont {Mele}},\
  and\ \bibinfo {author} {\bibfnamefont {A.}~\bibnamefont {Vishwanath}},\
  }\bibfield  {title} {\bibinfo {title} {{W}eyl and {D}irac semimetals in
  three-dimensional solids},\ }\href
  {https://doi.org/10.1103/RevModPhys.90.015001} {\bibfield  {journal}
  {\bibinfo  {journal} {Reviews of Modern Physics}\ }\textbf {\bibinfo {volume}
  {90}},\ \bibinfo {pages} {015001} (\bibinfo {year} {2018})}\BibitemShut
  {NoStop}%
\bibitem [{\citenamefont {M{\"u}hlbauer}\ \emph {et~al.}(2009)\citenamefont
  {M{\"u}hlbauer}, \citenamefont {Binz}, \citenamefont {Jonietz}, \citenamefont
  {Pfleiderer}, \citenamefont {Rosch}, \citenamefont {Neubauer}, \citenamefont
  {Georgii},\ and\ \citenamefont {B{\"o}ni}}]{Muhlbauer2009}%
  \BibitemOpen
  \bibfield  {author} {\bibinfo {author} {\bibfnamefont {S.}~\bibnamefont
  {M{\"u}hlbauer}}, \bibinfo {author} {\bibfnamefont {B.}~\bibnamefont {Binz}},
  \bibinfo {author} {\bibfnamefont {F.}~\bibnamefont {Jonietz}}, \bibinfo
  {author} {\bibfnamefont {C.}~\bibnamefont {Pfleiderer}}, \bibinfo {author}
  {\bibfnamefont {A.}~\bibnamefont {Rosch}}, \bibinfo {author} {\bibfnamefont
  {A.}~\bibnamefont {Neubauer}}, \bibinfo {author} {\bibfnamefont
  {R.}~\bibnamefont {Georgii}},\ and\ \bibinfo {author} {\bibfnamefont
  {P.}~\bibnamefont {B{\"o}ni}},\ }\bibfield  {title} {\bibinfo {title}
  {Skyrmion lattice in a chiral magnet},\ }\href
  {https://doi.org/10.1126/science.1166767} {\bibfield  {journal} {\bibinfo
  {journal} {Science}\ }\textbf {\bibinfo {volume} {323}},\ \bibinfo {pages}
  {915} (\bibinfo {year} {2009})}\BibitemShut {NoStop}%
\bibitem [{\citenamefont {Birch}\ \emph {et~al.}(2020)\citenamefont {Birch},
  \citenamefont {Cort{\'e}s-Ortu{\~n}o}, \citenamefont {Turnbull},
  \citenamefont {Wilson}, \citenamefont {Gro{\ss}}, \citenamefont {Tr{\"a}ger},
  \citenamefont {Laurenson}, \citenamefont {Bukin}, \citenamefont {Moody},
  \citenamefont {Weigand}, \citenamefont {Sch{\"u}tz}, \citenamefont {Popescu},
  \citenamefont {Fan}, \citenamefont {Steadman}, \citenamefont {Verezhak},
  \citenamefont {Balakrishnan}, \citenamefont {Loudon}, \citenamefont
  {Twitchett-Harrison}, \citenamefont {Hovorka}, \citenamefont {Fangohr},
  \citenamefont {Ogrin}, \citenamefont {Gr{\"a}fe},\ and\ \citenamefont
  {Hatton}}]{Birch2020}%
  \BibitemOpen
  \bibfield  {author} {\bibinfo {author} {\bibfnamefont {M.~T.}\ \bibnamefont
  {Birch}}, \bibinfo {author} {\bibfnamefont {D.}~\bibnamefont
  {Cort{\'e}s-Ortu{\~n}o}}, \bibinfo {author} {\bibfnamefont {L.~A.}\
  \bibnamefont {Turnbull}}, \bibinfo {author} {\bibfnamefont {M.~N.}\
  \bibnamefont {Wilson}}, \bibinfo {author} {\bibfnamefont {F.}~\bibnamefont
  {Gro{\ss}}}, \bibinfo {author} {\bibfnamefont {N.}~\bibnamefont
  {Tr{\"a}ger}}, \bibinfo {author} {\bibfnamefont {A.}~\bibnamefont
  {Laurenson}}, \bibinfo {author} {\bibfnamefont {N.}~\bibnamefont {Bukin}},
  \bibinfo {author} {\bibfnamefont {S.~H.}\ \bibnamefont {Moody}}, \bibinfo
  {author} {\bibfnamefont {M.}~\bibnamefont {Weigand}}, \bibinfo {author}
  {\bibfnamefont {G.}~\bibnamefont {Sch{\"u}tz}}, \bibinfo {author}
  {\bibfnamefont {H.}~\bibnamefont {Popescu}}, \bibinfo {author} {\bibfnamefont
  {R.}~\bibnamefont {Fan}}, \bibinfo {author} {\bibfnamefont {P.}~\bibnamefont
  {Steadman}}, \bibinfo {author} {\bibfnamefont {J.~A.~T.}\ \bibnamefont
  {Verezhak}}, \bibinfo {author} {\bibfnamefont {G.}~\bibnamefont
  {Balakrishnan}}, \bibinfo {author} {\bibfnamefont {J.~C.}\ \bibnamefont
  {Loudon}}, \bibinfo {author} {\bibfnamefont {A.~C.}\ \bibnamefont
  {Twitchett-Harrison}}, \bibinfo {author} {\bibfnamefont {O.}~\bibnamefont
  {Hovorka}}, \bibinfo {author} {\bibfnamefont {H.}~\bibnamefont {Fangohr}},
  \bibinfo {author} {\bibfnamefont {F.~Y.}\ \bibnamefont {Ogrin}}, \bibinfo
  {author} {\bibfnamefont {J.}~\bibnamefont {Gr{\"a}fe}},\ and\ \bibinfo
  {author} {\bibfnamefont {P.~D.}\ \bibnamefont {Hatton}},\ }\bibfield  {title}
  {\bibinfo {title} {Real-space imaging of confined magnetic skyrmion tubes},\
  }\href {https://doi.org/10.1038/s41467-020-15474-8} {\bibfield  {journal}
  {\bibinfo  {journal} {Nature Communications}\ }\textbf {\bibinfo {volume}
  {11}},\ \bibinfo {pages} {1726} (\bibinfo {year} {2020})}\BibitemShut
  {NoStop}%
\bibitem [{\citenamefont {Kurumaji}\ \emph {et~al.}(2019)\citenamefont
  {Kurumaji}, \citenamefont {Nakajima}, \citenamefont {Hirschberger},
  \citenamefont {Kikkawa}, \citenamefont {Yamasaki}, \citenamefont {Sagayama},
  \citenamefont {Nakao}, \citenamefont {Taguchi}, \citenamefont {Arima},\ and\
  \citenamefont {Tokura}}]{Kurumaji2019}%
  \BibitemOpen
  \bibfield  {author} {\bibinfo {author} {\bibfnamefont {T.}~\bibnamefont
  {Kurumaji}}, \bibinfo {author} {\bibfnamefont {T.}~\bibnamefont {Nakajima}},
  \bibinfo {author} {\bibfnamefont {M.}~\bibnamefont {Hirschberger}}, \bibinfo
  {author} {\bibfnamefont {A.}~\bibnamefont {Kikkawa}}, \bibinfo {author}
  {\bibfnamefont {Y.}~\bibnamefont {Yamasaki}}, \bibinfo {author}
  {\bibfnamefont {H.}~\bibnamefont {Sagayama}}, \bibinfo {author}
  {\bibfnamefont {H.}~\bibnamefont {Nakao}}, \bibinfo {author} {\bibfnamefont
  {Y.}~\bibnamefont {Taguchi}}, \bibinfo {author} {\bibfnamefont {T.-h.}\
  \bibnamefont {Arima}},\ and\ \bibinfo {author} {\bibfnamefont
  {Y.}~\bibnamefont {Tokura}},\ }\bibfield  {title} {\bibinfo {title} {Skyrmion
  lattice with a giant topological {H}all effect in a frustrated
  triangular-lattice magnet},\ }\href {https://doi.org/10.1126/science.aau0968}
  {\bibfield  {journal} {\bibinfo  {journal} {Science}\ }\textbf {\bibinfo
  {volume} {365}},\ \bibinfo {pages} {914} (\bibinfo {year}
  {2019})}\BibitemShut {NoStop}%
\bibitem [{\citenamefont {Jonietz}\ \emph {et~al.}(2010)\citenamefont
  {Jonietz}, \citenamefont {M{\"u}hlbauer}, \citenamefont {Pfleiderer},
  \citenamefont {Neubauer}, \citenamefont {M{\"u}nzer}, \citenamefont {Bauer},
  \citenamefont {Adams}, \citenamefont {Georgii}, \citenamefont {B{\"o}ni},
  \citenamefont {Duine}, \citenamefont {Everschor}, \citenamefont {Garst},\
  and\ \citenamefont {Rosch}}]{Jonietz2010}%
  \BibitemOpen
  \bibfield  {author} {\bibinfo {author} {\bibfnamefont {F.}~\bibnamefont
  {Jonietz}}, \bibinfo {author} {\bibfnamefont {S.}~\bibnamefont
  {M{\"u}hlbauer}}, \bibinfo {author} {\bibfnamefont {C.}~\bibnamefont
  {Pfleiderer}}, \bibinfo {author} {\bibfnamefont {A.}~\bibnamefont
  {Neubauer}}, \bibinfo {author} {\bibfnamefont {W.}~\bibnamefont
  {M{\"u}nzer}}, \bibinfo {author} {\bibfnamefont {A.}~\bibnamefont {Bauer}},
  \bibinfo {author} {\bibfnamefont {T.}~\bibnamefont {Adams}}, \bibinfo
  {author} {\bibfnamefont {R.}~\bibnamefont {Georgii}}, \bibinfo {author}
  {\bibfnamefont {P.}~\bibnamefont {B{\"o}ni}}, \bibinfo {author}
  {\bibfnamefont {R.~A.}\ \bibnamefont {Duine}}, \bibinfo {author}
  {\bibfnamefont {K.}~\bibnamefont {Everschor}}, \bibinfo {author}
  {\bibfnamefont {M.}~\bibnamefont {Garst}},\ and\ \bibinfo {author}
  {\bibfnamefont {A.}~\bibnamefont {Rosch}},\ }\bibfield  {title} {\bibinfo
  {title} {Spin transfer torques in \ce{MnSi} at ultralow current densities},\
  }\href {https://doi.org/10.1126/SCIENCE.1195709} {\bibfield  {journal}
  {\bibinfo  {journal} {Science}\ }\textbf {\bibinfo {volume} {330}},\ \bibinfo
  {pages} {1648} (\bibinfo {year} {2010})}\BibitemShut {NoStop}%
\bibitem [{\citenamefont {Peng}\ \emph {et~al.}(2021)\citenamefont {Peng},
  \citenamefont {Karube}, \citenamefont {Taguchi}, \citenamefont {Nagaosa},
  \citenamefont {Tokura},\ and\ \citenamefont {Yu}}]{Peng2021}%
  \BibitemOpen
  \bibfield  {author} {\bibinfo {author} {\bibfnamefont {L.}~\bibnamefont
  {Peng}}, \bibinfo {author} {\bibfnamefont {K.}~\bibnamefont {Karube}},
  \bibinfo {author} {\bibfnamefont {Y.}~\bibnamefont {Taguchi}}, \bibinfo
  {author} {\bibfnamefont {N.}~\bibnamefont {Nagaosa}}, \bibinfo {author}
  {\bibfnamefont {Y.}~\bibnamefont {Tokura}},\ and\ \bibinfo {author}
  {\bibfnamefont {X.}~\bibnamefont {Yu}},\ }\bibfield  {title} {\bibinfo
  {title} {Dynamic transition of current-driven single-skyrmion motion in a
  room-temperature chiral-lattice magnet},\ }\href
  {https://doi.org/10.1038/s41467-021-27073-2} {\bibfield  {journal} {\bibinfo
  {journal} {Nature Communications}\ }\textbf {\bibinfo {volume} {12}},\
  \bibinfo {pages} {6797} (\bibinfo {year} {2021})}\BibitemShut {NoStop}%
\bibitem [{\citenamefont {Luo}\ \emph {et~al.}(2020)\citenamefont {Luo},
  \citenamefont {Lin}, \citenamefont {Leroux}, \citenamefont {Wakeham},
  \citenamefont {Fobes}, \citenamefont {Bauer}, \citenamefont {Betts},
  \citenamefont {Thompson}, \citenamefont {Migliori}, \citenamefont
  {Janoschek},\ and\ \citenamefont {Maiorov}}]{Luo2020}%
  \BibitemOpen
  \bibfield  {author} {\bibinfo {author} {\bibfnamefont {Y.}~\bibnamefont
  {Luo}}, \bibinfo {author} {\bibfnamefont {S.-Z.}\ \bibnamefont {Lin}},
  \bibinfo {author} {\bibfnamefont {M.}~\bibnamefont {Leroux}}, \bibinfo
  {author} {\bibfnamefont {N.}~\bibnamefont {Wakeham}}, \bibinfo {author}
  {\bibfnamefont {D.~M.}\ \bibnamefont {Fobes}}, \bibinfo {author}
  {\bibfnamefont {E.~D.}\ \bibnamefont {Bauer}}, \bibinfo {author}
  {\bibfnamefont {J.~B.}\ \bibnamefont {Betts}}, \bibinfo {author}
  {\bibfnamefont {J.~D.}\ \bibnamefont {Thompson}}, \bibinfo {author}
  {\bibfnamefont {A.}~\bibnamefont {Migliori}}, \bibinfo {author}
  {\bibfnamefont {M.}~\bibnamefont {Janoschek}},\ and\ \bibinfo {author}
  {\bibfnamefont {B.}~\bibnamefont {Maiorov}},\ }\bibfield  {title} {\bibinfo
  {title} {Skyrmion lattice creep at ultra-low current densities},\ }\href
  {https://doi.org/10.1038/s43246-020-00083-1} {\bibfield  {journal} {\bibinfo
  {journal} {Communications Materials}\ }\textbf {\bibinfo {volume} {1}},\
  \bibinfo {pages} {83} (\bibinfo {year} {2020})}\BibitemShut {NoStop}%
\bibitem [{\citenamefont {Birch}\ \emph {et~al.}(2024)\citenamefont {Birch},
  \citenamefont {Belopolski}, \citenamefont {Fujishiro}, \citenamefont
  {Kawamura}, \citenamefont {Kikkawa}, \citenamefont {Taguchi}, \citenamefont
  {Hirschberger}, \citenamefont {Nagaosa},\ and\ \citenamefont
  {Tokura}}]{Birch2024}%
  \BibitemOpen
  \bibfield  {author} {\bibinfo {author} {\bibfnamefont {M.~T.}\ \bibnamefont
  {Birch}}, \bibinfo {author} {\bibfnamefont {I.}~\bibnamefont {Belopolski}},
  \bibinfo {author} {\bibfnamefont {Y.}~\bibnamefont {Fujishiro}}, \bibinfo
  {author} {\bibfnamefont {M.}~\bibnamefont {Kawamura}}, \bibinfo {author}
  {\bibfnamefont {A.}~\bibnamefont {Kikkawa}}, \bibinfo {author} {\bibfnamefont
  {Y.}~\bibnamefont {Taguchi}}, \bibinfo {author} {\bibfnamefont
  {M.}~\bibnamefont {Hirschberger}}, \bibinfo {author} {\bibfnamefont
  {N.}~\bibnamefont {Nagaosa}},\ and\ \bibinfo {author} {\bibfnamefont
  {Y.}~\bibnamefont {Tokura}},\ }\bibfield  {title} {\bibinfo {title} {Dynamic
  transition and galilean relativity of current-driven skyrmions},\ }\href
  {https://doi.org/10.1038/s41586-024-07859-2} {\bibfield  {journal} {\bibinfo
  {journal} {Nature}\ }\textbf {\bibinfo {volume} {633}},\ \bibinfo {pages}
  {554} (\bibinfo {year} {2024})}\BibitemShut {NoStop}%
\bibitem [{\citenamefont {B{\"u}ttner}\ \emph {et~al.}(2015)\citenamefont
  {B{\"u}ttner}, \citenamefont {Moutafis}, \citenamefont {Schneider},
  \citenamefont {Kr{\"u}ger}, \citenamefont {G{\"u}nther}, \citenamefont
  {Geilhufe}, \citenamefont {Schmising}, \citenamefont {Mohanty}, \citenamefont
  {Pfau}, \citenamefont {Schaffert}, \citenamefont {Bisig}, \citenamefont
  {Foerster}, \citenamefont {Schulz}, \citenamefont {Vaz}, \citenamefont
  {Franken}, \citenamefont {Swagten}, \citenamefont {Kl{\"a}ui},\ and\
  \citenamefont {Eisebitt}}]{Buttner2015}%
  \BibitemOpen
  \bibfield  {author} {\bibinfo {author} {\bibfnamefont {F.}~\bibnamefont
  {B{\"u}ttner}}, \bibinfo {author} {\bibfnamefont {C.}~\bibnamefont
  {Moutafis}}, \bibinfo {author} {\bibfnamefont {M.}~\bibnamefont {Schneider}},
  \bibinfo {author} {\bibfnamefont {B.}~\bibnamefont {Kr{\"u}ger}}, \bibinfo
  {author} {\bibfnamefont {C.~M.}\ \bibnamefont {G{\"u}nther}}, \bibinfo
  {author} {\bibfnamefont {J.}~\bibnamefont {Geilhufe}}, \bibinfo {author}
  {\bibfnamefont {C.~v.~K.}\ \bibnamefont {Schmising}}, \bibinfo {author}
  {\bibfnamefont {J.}~\bibnamefont {Mohanty}}, \bibinfo {author} {\bibfnamefont
  {B.}~\bibnamefont {Pfau}}, \bibinfo {author} {\bibfnamefont {S.}~\bibnamefont
  {Schaffert}}, \bibinfo {author} {\bibfnamefont {A.}~\bibnamefont {Bisig}},
  \bibinfo {author} {\bibfnamefont {M.}~\bibnamefont {Foerster}}, \bibinfo
  {author} {\bibfnamefont {T.}~\bibnamefont {Schulz}}, \bibinfo {author}
  {\bibfnamefont {C.~A.~F.}\ \bibnamefont {Vaz}}, \bibinfo {author}
  {\bibfnamefont {J.~H.}\ \bibnamefont {Franken}}, \bibinfo {author}
  {\bibfnamefont {H.~J.~M.}\ \bibnamefont {Swagten}}, \bibinfo {author}
  {\bibfnamefont {M.}~\bibnamefont {Kl{\"a}ui}},\ and\ \bibinfo {author}
  {\bibfnamefont {S.}~\bibnamefont {Eisebitt}},\ }\bibfield  {title} {\bibinfo
  {title} {Dynamics and inertia of skyrmionic spin structures},\ }\href
  {https://doi.org/10.1038/nphys3234} {\bibfield  {journal} {\bibinfo
  {journal} {Nature Physics}\ }\textbf {\bibinfo {volume} {11}},\ \bibinfo
  {pages} {225} (\bibinfo {year} {2015})}\BibitemShut {NoStop}%
\bibitem [{\citenamefont {Reichhardt}\ \emph {et~al.}(2015)\citenamefont
  {Reichhardt}, \citenamefont {Ray},\ and\ \citenamefont
  {Reichhardt}}]{Reichhardt2015}%
  \BibitemOpen
  \bibfield  {author} {\bibinfo {author} {\bibfnamefont {C.}~\bibnamefont
  {Reichhardt}}, \bibinfo {author} {\bibfnamefont {D.}~\bibnamefont {Ray}},\
  and\ \bibinfo {author} {\bibfnamefont {C.~J.~O.}\ \bibnamefont
  {Reichhardt}},\ }\bibfield  {title} {\bibinfo {title} {Collective transport
  properties of driven skyrmions with random disorder},\ }\href
  {https://doi.org/10.1103/PhysRevLett.114.217202} {\bibfield  {journal}
  {\bibinfo  {journal} {Physical Review Letters}\ }\textbf {\bibinfo {volume}
  {114}},\ \bibinfo {pages} {217202} (\bibinfo {year} {2015})}\BibitemShut
  {NoStop}%
\bibitem [{\citenamefont {Yokouchi}\ \emph {et~al.}(2017)\citenamefont
  {Yokouchi}, \citenamefont {Kanazawa}, \citenamefont {Kikkawa}, \citenamefont
  {Morikawa}, \citenamefont {Shibata}, \citenamefont {Arima}, \citenamefont
  {Taguchi}, \citenamefont {Kagawa},\ and\ \citenamefont
  {Tokura}}]{Yokouchi2017}%
  \BibitemOpen
  \bibfield  {author} {\bibinfo {author} {\bibfnamefont {T.}~\bibnamefont
  {Yokouchi}}, \bibinfo {author} {\bibfnamefont {N.}~\bibnamefont {Kanazawa}},
  \bibinfo {author} {\bibfnamefont {A.}~\bibnamefont {Kikkawa}}, \bibinfo
  {author} {\bibfnamefont {D.}~\bibnamefont {Morikawa}}, \bibinfo {author}
  {\bibfnamefont {K.}~\bibnamefont {Shibata}}, \bibinfo {author} {\bibfnamefont
  {T.}~\bibnamefont {Arima}}, \bibinfo {author} {\bibfnamefont
  {Y.}~\bibnamefont {Taguchi}}, \bibinfo {author} {\bibfnamefont
  {F.}~\bibnamefont {Kagawa}},\ and\ \bibinfo {author} {\bibfnamefont
  {Y.}~\bibnamefont {Tokura}},\ }\bibfield  {title} {\bibinfo {title}
  {Electrical magnetochiral effect induced by chiral spin fluctuations},\
  }\href {https://doi.org/10.1038/s41467-017-01094-2} {\bibfield  {journal}
  {\bibinfo  {journal} {Nature Communications}\ }\textbf {\bibinfo {volume}
  {8}},\ \bibinfo {pages} {866} (\bibinfo {year} {2017})}\BibitemShut {NoStop}%
\bibitem [{\citenamefont {Schulz}\ \emph {et~al.}(2012)\citenamefont {Schulz},
  \citenamefont {Ritz}, \citenamefont {Bauer}, \citenamefont {Halder},
  \citenamefont {Wagner}, \citenamefont {Franz}, \citenamefont {Pfleiderer},
  \citenamefont {Everschor}, \citenamefont {Garst},\ and\ \citenamefont
  {Rosch}}]{Schulz2012}%
  \BibitemOpen
  \bibfield  {author} {\bibinfo {author} {\bibfnamefont {T.}~\bibnamefont
  {Schulz}}, \bibinfo {author} {\bibfnamefont {R.}~\bibnamefont {Ritz}},
  \bibinfo {author} {\bibfnamefont {A.}~\bibnamefont {Bauer}}, \bibinfo
  {author} {\bibfnamefont {M.}~\bibnamefont {Halder}}, \bibinfo {author}
  {\bibfnamefont {M.}~\bibnamefont {Wagner}}, \bibinfo {author} {\bibfnamefont
  {C.}~\bibnamefont {Franz}}, \bibinfo {author} {\bibfnamefont
  {C.}~\bibnamefont {Pfleiderer}}, \bibinfo {author} {\bibfnamefont
  {K.}~\bibnamefont {Everschor}}, \bibinfo {author} {\bibfnamefont
  {M.}~\bibnamefont {Garst}},\ and\ \bibinfo {author} {\bibfnamefont
  {A.}~\bibnamefont {Rosch}},\ }\bibfield  {title} {\bibinfo {title} {Emergent
  electrodynamics of skyrmions in a chiral magnet},\ }\href
  {https://doi.org/10.1038/nphys2231} {\bibfield  {journal} {\bibinfo
  {journal} {Nature Physics}\ }\textbf {\bibinfo {volume} {8}},\ \bibinfo
  {pages} {301} (\bibinfo {year} {2012})}\BibitemShut {NoStop}%
\bibitem [{\citenamefont {Volovik}(1987)}]{Volovik1987}%
  \BibitemOpen
  \bibfield  {author} {\bibinfo {author} {\bibfnamefont {G.~E.}\ \bibnamefont
  {Volovik}},\ }\bibfield  {title} {\bibinfo {title} {Linear momentum in
  ferromagnets},\ }\href {https://doi.org/10.1088/0022-3719/20/7/003}
  {\bibfield  {journal} {\bibinfo  {journal} {Journal of Physics C: Solid State
  Physics}\ }\textbf {\bibinfo {volume} {20}},\ \bibinfo {pages} {L83}
  (\bibinfo {year} {1987})}\BibitemShut {NoStop}%
\bibitem [{\citenamefont {Nagaosa}\ and\ \citenamefont
  {Tokura}(2013)}]{Nagaosa2013}%
  \BibitemOpen
  \bibfield  {author} {\bibinfo {author} {\bibfnamefont {N.}~\bibnamefont
  {Nagaosa}}\ and\ \bibinfo {author} {\bibfnamefont {Y.}~\bibnamefont
  {Tokura}},\ }\bibfield  {title} {\bibinfo {title} {Topological properties and
  dynamics of magnetic skyrmions},\ }\href
  {https://doi.org/10.1038/nnano.2013.243} {\bibfield  {journal} {\bibinfo
  {journal} {Nature Nanotechnology}\ }\textbf {\bibinfo {volume} {8}},\
  \bibinfo {pages} {899} (\bibinfo {year} {2013})}\BibitemShut {NoStop}%
\bibitem [{\citenamefont {Zang}\ \emph {et~al.}(2011)\citenamefont {Zang},
  \citenamefont {Mostovoy}, \citenamefont {Han},\ and\ \citenamefont
  {Nagaosa}}]{Zang2011}%
  \BibitemOpen
  \bibfield  {author} {\bibinfo {author} {\bibfnamefont {J.}~\bibnamefont
  {Zang}}, \bibinfo {author} {\bibfnamefont {M.}~\bibnamefont {Mostovoy}},
  \bibinfo {author} {\bibfnamefont {J.~H.}\ \bibnamefont {Han}},\ and\ \bibinfo
  {author} {\bibfnamefont {N.}~\bibnamefont {Nagaosa}},\ }\bibfield  {title}
  {\bibinfo {title} {Dynamics of skyrmion crystals in metallic thin films},\
  }\href {https://doi.org/10.1103/PhysRevLett.107.136804} {\bibfield  {journal}
  {\bibinfo  {journal} {Physical Review Letters}\ }\textbf {\bibinfo {volume}
  {107}},\ \bibinfo {pages} {136804} (\bibinfo {year} {2011})}\BibitemShut
  {NoStop}%
\bibitem [{\citenamefont {Nagaosa}(2019)}]{Nagaosa2019}%
  \BibitemOpen
  \bibfield  {author} {\bibinfo {author} {\bibfnamefont {N.}~\bibnamefont
  {Nagaosa}},\ }\bibfield  {title} {\bibinfo {title} {Emergent inductor by
  spiral magnets},\ }\href {https://doi.org/10.7567/1347-4065/ab5294}
  {\bibfield  {journal} {\bibinfo  {journal} {Japanese Journal of Applied
  Physics}\ }\textbf {\bibinfo {volume} {58}},\ \bibinfo {pages} {120909}
  (\bibinfo {year} {2019})}\BibitemShut {NoStop}%
\bibitem [{\citenamefont {Yokouchi}\ \emph {et~al.}(2020)\citenamefont
  {Yokouchi}, \citenamefont {Kagawa}, \citenamefont {Hirschberger},
  \citenamefont {Otani}, \citenamefont {Nagaosa},\ and\ \citenamefont
  {Tokura}}]{Yokouchi2020}%
  \BibitemOpen
  \bibfield  {author} {\bibinfo {author} {\bibfnamefont {T.}~\bibnamefont
  {Yokouchi}}, \bibinfo {author} {\bibfnamefont {F.}~\bibnamefont {Kagawa}},
  \bibinfo {author} {\bibfnamefont {M.}~\bibnamefont {Hirschberger}}, \bibinfo
  {author} {\bibfnamefont {Y.}~\bibnamefont {Otani}}, \bibinfo {author}
  {\bibfnamefont {N.}~\bibnamefont {Nagaosa}},\ and\ \bibinfo {author}
  {\bibfnamefont {Y.}~\bibnamefont {Tokura}},\ }\bibfield  {title} {\bibinfo
  {title} {Emergent electromagnetic induction in a helical-spin magnet},\
  }\href {https://doi.org/10.1038/s41586-020-2775-x} {\bibfield  {journal}
  {\bibinfo  {journal} {Nature}\ }\textbf {\bibinfo {volume} {586}},\ \bibinfo
  {pages} {232} (\bibinfo {year} {2020})}\BibitemShut {NoStop}%
\bibitem [{\citenamefont {Kitaori}\ \emph {et~al.}(2021)\citenamefont
  {Kitaori}, \citenamefont {Kanazawa}, \citenamefont {Yokouchi}, \citenamefont
  {Kagawa}, \citenamefont {Nagaosa},\ and\ \citenamefont
  {Tokura}}]{Kitaori2021}%
  \BibitemOpen
  \bibfield  {author} {\bibinfo {author} {\bibfnamefont {A.}~\bibnamefont
  {Kitaori}}, \bibinfo {author} {\bibfnamefont {N.}~\bibnamefont {Kanazawa}},
  \bibinfo {author} {\bibfnamefont {T.}~\bibnamefont {Yokouchi}}, \bibinfo
  {author} {\bibfnamefont {F.}~\bibnamefont {Kagawa}}, \bibinfo {author}
  {\bibfnamefont {N.}~\bibnamefont {Nagaosa}},\ and\ \bibinfo {author}
  {\bibfnamefont {Y.}~\bibnamefont {Tokura}},\ }\bibfield  {title} {\bibinfo
  {title} {Emergent electromagnetic induction beyond room temperature},\ }\href
  {https://doi.org/10.1073/pnas.2105422118} {\bibfield  {journal} {\bibinfo
  {journal} {Proceedings of the National Academy of Sciences}\ }\textbf
  {\bibinfo {volume} {118}},\ \bibinfo {pages} {e2105422118} (\bibinfo {year}
  {2021})}\BibitemShut {NoStop}%
\bibitem [{\citenamefont {Zhang}\ \emph {et~al.}(2023)\citenamefont {Zhang},
  \citenamefont {Matsushima}, \citenamefont {Ohashi}, \citenamefont
  {Matsuzaka},\ and\ \citenamefont {Kaiju}}]{Zhang2023}%
  \BibitemOpen
  \bibfield  {author} {\bibinfo {author} {\bibfnamefont {Z.}~\bibnamefont
  {Zhang}}, \bibinfo {author} {\bibfnamefont {Y.}~\bibnamefont {Matsushima}},
  \bibinfo {author} {\bibfnamefont {Y.}~\bibnamefont {Ohashi}}, \bibinfo
  {author} {\bibfnamefont {M.}~\bibnamefont {Matsuzaka}},\ and\ \bibinfo
  {author} {\bibfnamefont {H.}~\bibnamefont {Kaiju}},\ }\bibfield  {title}
  {\bibinfo {title} {Emergent magneto-inductance effect in \ce{Ni45Fe55} thin
  films on polycarbonate substrates},\ }\href
  {https://doi.org/10.1109/INTERMAGShortPapers58606.2023.10228840} {\bibfield
  {journal} {\bibinfo  {journal} {2023 IEEE International Magnetic Conference -
  Short Papers (INTERMAG Short Papers)}\ ,\ \bibinfo {pages} {1}} (\bibinfo
  {year} {2023})}\BibitemShut {NoStop}%
\bibitem [{\citenamefont {Kitaori}\ \emph {et~al.}(2024)\citenamefont
  {Kitaori}, \citenamefont {White}, \citenamefont {Ukleev}, \citenamefont
  {Peng}, \citenamefont {Nakajima}, \citenamefont {Kanazawa}, \citenamefont
  {Yu}, \citenamefont {{\=O}nuki},\ and\ \citenamefont {Tokura}}]{Kitaori2024}%
  \BibitemOpen
  \bibfield  {author} {\bibinfo {author} {\bibfnamefont {A.}~\bibnamefont
  {Kitaori}}, \bibinfo {author} {\bibfnamefont {J.~S.}\ \bibnamefont {White}},
  \bibinfo {author} {\bibfnamefont {V.}~\bibnamefont {Ukleev}}, \bibinfo
  {author} {\bibfnamefont {L.}~\bibnamefont {Peng}}, \bibinfo {author}
  {\bibfnamefont {K.}~\bibnamefont {Nakajima}}, \bibinfo {author}
  {\bibfnamefont {N.}~\bibnamefont {Kanazawa}}, \bibinfo {author}
  {\bibfnamefont {X.}~\bibnamefont {Yu}}, \bibinfo {author} {\bibfnamefont
  {Y.}~\bibnamefont {{\=O}nuki}},\ and\ \bibinfo {author} {\bibfnamefont
  {Y.}~\bibnamefont {Tokura}},\ }\bibfield  {title} {\bibinfo {title} {Enhanced
  emergent electromagnetic inductance in {{\ce{Tb5Sb3}}} due to highly
  disordered helimagnetism},\ }\href
  {https://doi.org/10.1038/s42005-024-01656-z} {\bibfield  {journal} {\bibinfo
  {journal} {Communications Physics}\ }\textbf {\bibinfo {volume} {7}},\
  \bibinfo {pages} {159} (\bibinfo {year} {2024})}\BibitemShut {NoStop}%
\bibitem [{\citenamefont {Kurebayashi}\ and\ \citenamefont
  {Nagaosa}(2021)}]{Kurebayashi2021}%
  \BibitemOpen
  \bibfield  {author} {\bibinfo {author} {\bibfnamefont {D.}~\bibnamefont
  {Kurebayashi}}\ and\ \bibinfo {author} {\bibfnamefont {N.}~\bibnamefont
  {Nagaosa}},\ }\bibfield  {title} {\bibinfo {title} {Electromagnetic response
  in spiral magnets and emergent inductance},\ }\href
  {https://doi.org/10.1038/s42005-021-00765-3} {\bibfield  {journal} {\bibinfo
  {journal} {Communications Physics}\ }\textbf {\bibinfo {volume} {4}},\
  \bibinfo {pages} {260} (\bibinfo {year} {2021})}\BibitemShut {NoStop}%
\bibitem [{\citenamefont {Ieda}\ and\ \citenamefont {Yamane}(2021)}]{Ieda2021}%
  \BibitemOpen
  \bibfield  {author} {\bibinfo {author} {\bibfnamefont {J.}~\bibnamefont
  {Ieda}}\ and\ \bibinfo {author} {\bibfnamefont {Y.}~\bibnamefont {Yamane}},\
  }\bibfield  {title} {\bibinfo {title} {Intrinsic and extrinsic tunability of
  rashba spin-orbit coupled emergent inductors},\ }\href
  {https://doi.org/10.1103/PhysRevB.103.L100402} {\bibfield  {journal}
  {\bibinfo  {journal} {Physical Review B}\ }\textbf {\bibinfo {volume}
  {103}},\ \bibinfo {pages} {L100402} (\bibinfo {year} {2021})}\BibitemShut
  {NoStop}%
\bibitem [{\citenamefont {Yamane}\ \emph {et~al.}(2022)\citenamefont {Yamane},
  \citenamefont {Fukami},\ and\ \citenamefont {Ieda}}]{Yamane2022}%
  \BibitemOpen
  \bibfield  {author} {\bibinfo {author} {\bibfnamefont {Y.}~\bibnamefont
  {Yamane}}, \bibinfo {author} {\bibfnamefont {S.}~\bibnamefont {Fukami}},\
  and\ \bibinfo {author} {\bibfnamefont {J.}~\bibnamefont {Ieda}},\ }\bibfield
  {title} {\bibinfo {title} {Theory of emergent inductance with spin-orbit
  coupling effects},\ }\href {https://doi.org/10.1103/PhysRevLett.128.147201}
  {\bibfield  {journal} {\bibinfo  {journal} {Physical Review Letters}\
  }\textbf {\bibinfo {volume} {128}},\ \bibinfo {pages} {147201} (\bibinfo
  {year} {2022})}\BibitemShut {NoStop}%
\bibitem [{\citenamefont {Shimizu}\ \emph {et~al.}(2023)\citenamefont
  {Shimizu}, \citenamefont {Okumura}, \citenamefont {Kato},\ and\ \citenamefont
  {Motome}}]{Shimizu2023}%
  \BibitemOpen
  \bibfield  {author} {\bibinfo {author} {\bibfnamefont {K.}~\bibnamefont
  {Shimizu}}, \bibinfo {author} {\bibfnamefont {S.}~\bibnamefont {Okumura}},
  \bibinfo {author} {\bibfnamefont {Y.}~\bibnamefont {Kato}},\ and\ \bibinfo
  {author} {\bibfnamefont {Y.}~\bibnamefont {Motome}},\ }\bibfield  {title}
  {\bibinfo {title} {Emergent electric field from magnetic resonances in a
  one-dimensional chiral magnet},\ }\href
  {https://doi.org/10.1103/PhysRevB.108.134436} {\bibfield  {journal} {\bibinfo
   {journal} {Physical Review B}\ }\textbf {\bibinfo {volume} {108}},\ \bibinfo
  {pages} {134436} (\bibinfo {year} {2023})}\BibitemShut {NoStop}%
\bibitem [{\citenamefont {Araki}\ and\ \citenamefont {Ieda}(2023)}]{Araki2023}%
  \BibitemOpen
  \bibfield  {author} {\bibinfo {author} {\bibfnamefont {Y.}~\bibnamefont
  {Araki}}\ and\ \bibinfo {author} {\bibfnamefont {J.}~\bibnamefont {Ieda}},\
  }\bibfield  {title} {\bibinfo {title} {Emergence of inductance and
  capacitance from topological electromagnetism},\ }\href
  {https://doi.org/10.7566/JPSJ.92.074705} {\bibfield  {journal} {\bibinfo
  {journal} {Journal of the Physical Society of Japan}\ }\textbf {\bibinfo
  {volume} {92}},\ \bibinfo {pages} {074705} (\bibinfo {year}
  {2023})}\BibitemShut {NoStop}%
\bibitem [{\citenamefont {Oh}\ and\ \citenamefont {Nagaosa}(2024)}]{Oh2024}%
  \BibitemOpen
  \bibfield  {author} {\bibinfo {author} {\bibfnamefont {T.}~\bibnamefont
  {Oh}}\ and\ \bibinfo {author} {\bibfnamefont {N.}~\bibnamefont {Nagaosa}},\
  }\bibfield  {title} {\bibinfo {title} {Emergent inductance from spin
  fluctuations in strongly correlated magnets},\ }\href
  {https://doi.org/10.1103/PhysRevLett.132.116501} {\bibfield  {journal}
  {\bibinfo  {journal} {Physical Review Letters}\ }\textbf {\bibinfo {volume}
  {132}},\ \bibinfo {pages} {116501} (\bibinfo {year} {2024})}\BibitemShut
  {NoStop}%
\bibitem [{\citenamefont {Furuta}\ \emph {et~al.}(2023)\citenamefont {Furuta},
  \citenamefont {Koshibae},\ and\ \citenamefont {Kagawa}}]{Furuta2023}%
  \BibitemOpen
  \bibfield  {author} {\bibinfo {author} {\bibfnamefont {S.}~\bibnamefont
  {Furuta}}, \bibinfo {author} {\bibfnamefont {W.}~\bibnamefont {Koshibae}},\
  and\ \bibinfo {author} {\bibfnamefont {F.}~\bibnamefont {Kagawa}},\
  }\bibfield  {title} {\bibinfo {title} {Symmetry of the emergent inductance
  tensor exhibited by magnetic textures},\ }\href
  {https://doi.org/10.1038/s44306-023-00001-4} {\bibfield  {journal} {\bibinfo
  {journal} {npj Spintronics}\ }\textbf {\bibinfo {volume} {1}},\ \bibinfo
  {pages} {3} (\bibinfo {year} {2023})}\BibitemShut {NoStop}%
\bibitem [{\citenamefont {Scheuchenpflug}\ \emph {et~al.}(2025)\citenamefont
  {Scheuchenpflug}, \citenamefont {Esser}, \citenamefont {Gruhl}, \citenamefont
  {Hirschberger},\ and\ \citenamefont {Gegenwart}}]{Scheuchenpflug2025}%
  \BibitemOpen
  \bibfield  {author} {\bibinfo {author} {\bibfnamefont {L.}~\bibnamefont
  {Scheuchenpflug}}, \bibinfo {author} {\bibfnamefont {S.}~\bibnamefont
  {Esser}}, \bibinfo {author} {\bibfnamefont {R.}~\bibnamefont {Gruhl}},
  \bibinfo {author} {\bibfnamefont {M.}~\bibnamefont {Hirschberger}},\ and\
  \bibinfo {author} {\bibfnamefont {P.}~\bibnamefont {Gegenwart}},\ }\href
  {https://arxiv.org/abs/2503.10600} {\bibinfo {title} {Current-linear emergent
  induction of pinned skyrmion textures in an oxide bilayer}} (\bibinfo {year}
  {2025}),\ \Eprint {https://arxiv.org/abs/2503.10600} {arXiv:2503.10600
  [cond-mat.mes-hall]} \BibitemShut {NoStop}%
\bibitem [{\citenamefont {Yamanouchi}\ \emph {et~al.}(2004)\citenamefont
  {Yamanouchi}, \citenamefont {Chiba}, \citenamefont {Matsukura},\ and\
  \citenamefont {Ohno}}]{Yamanouchi2004}%
  \BibitemOpen
  \bibfield  {author} {\bibinfo {author} {\bibfnamefont {M.}~\bibnamefont
  {Yamanouchi}}, \bibinfo {author} {\bibfnamefont {D.}~\bibnamefont {Chiba}},
  \bibinfo {author} {\bibfnamefont {F.}~\bibnamefont {Matsukura}},\ and\
  \bibinfo {author} {\bibfnamefont {H.}~\bibnamefont {Ohno}},\ }\bibfield
  {title} {\bibinfo {title} {Current-induced domain-wall switching in a
  ferromagnetic semiconductor structure},\ }\href
  {https://doi.org/10.1038/nature02441} {\bibfield  {journal} {\bibinfo
  {journal} {Nature}\ }\textbf {\bibinfo {volume} {428}},\ \bibinfo {pages}
  {539} (\bibinfo {year} {2004})}\BibitemShut {NoStop}%
\bibitem [{\citenamefont {Kimoto}\ \emph {et~al.}(2025)\citenamefont {Kimoto},
  \citenamefont {Masuda}, \citenamefont {Seki}, \citenamefont {Nii},
  \citenamefont {Ohe}, \citenamefont {Nambu},\ and\ \citenamefont
  {Onose}}]{Kimoto2025}%
  \BibitemOpen
  \bibfield  {author} {\bibinfo {author} {\bibfnamefont {Y.}~\bibnamefont
  {Kimoto}}, \bibinfo {author} {\bibfnamefont {H.}~\bibnamefont {Masuda}},
  \bibinfo {author} {\bibfnamefont {T.}~\bibnamefont {Seki}}, \bibinfo {author}
  {\bibfnamefont {Y.}~\bibnamefont {Nii}}, \bibinfo {author} {\bibfnamefont
  {J.-i.}\ \bibnamefont {Ohe}}, \bibinfo {author} {\bibfnamefont
  {Y.}~\bibnamefont {Nambu}},\ and\ \bibinfo {author} {\bibfnamefont
  {Y.}~\bibnamefont {Onose}},\ }\bibfield  {title} {\bibinfo {title}
  {Current-induced sliding motion in a helimagnet \ce{MnAu2}},\ }\href
  {https://doi.org/10.1103/PhysRevLett.134.056702} {\bibfield  {journal}
  {\bibinfo  {journal} {Physical Review Letters}\ }\textbf {\bibinfo {volume}
  {134}},\ \bibinfo {pages} {056702} (\bibinfo {year} {2025})}\BibitemShut
  {NoStop}%
\bibitem [{\citenamefont {Neubauer}\ \emph {et~al.}(2009)\citenamefont
  {Neubauer}, \citenamefont {Pfleiderer}, \citenamefont {Binz}, \citenamefont
  {Rosch}, \citenamefont {Ritz}, \citenamefont {Niklowitz},\ and\ \citenamefont
  {B{\"o}ni}}]{Neubauer2009}%
  \BibitemOpen
  \bibfield  {author} {\bibinfo {author} {\bibfnamefont {A.}~\bibnamefont
  {Neubauer}}, \bibinfo {author} {\bibfnamefont {C.}~\bibnamefont
  {Pfleiderer}}, \bibinfo {author} {\bibfnamefont {B.}~\bibnamefont {Binz}},
  \bibinfo {author} {\bibfnamefont {A.}~\bibnamefont {Rosch}}, \bibinfo
  {author} {\bibfnamefont {R.}~\bibnamefont {Ritz}}, \bibinfo {author}
  {\bibfnamefont {P.~G.}\ \bibnamefont {Niklowitz}},\ and\ \bibinfo {author}
  {\bibfnamefont {P.}~\bibnamefont {B{\"o}ni}},\ }\bibfield  {title} {\bibinfo
  {title} {Topological {H}all effect in the {A} phase of \ce{MnSi}},\ }\href
  {https://doi.org/10.1103/PhysRevLett.102.186602} {\bibfield  {journal}
  {\bibinfo  {journal} {Physical Review Letters}\ }\textbf {\bibinfo {volume}
  {102}} (\bibinfo {year} {2009})}\BibitemShut {NoStop}%
\bibitem [{\citenamefont {Sato}\ \emph {et~al.}(2022)\citenamefont {Sato},
  \citenamefont {Koshibae}, \citenamefont {Kikkawa}, \citenamefont {Taguchi},
  \citenamefont {Nagaosa}, \citenamefont {Tokura},\ and\ \citenamefont
  {Kagawa}}]{Sato2022}%
  \BibitemOpen
  \bibfield  {author} {\bibinfo {author} {\bibfnamefont {T.}~\bibnamefont
  {Sato}}, \bibinfo {author} {\bibfnamefont {W.}~\bibnamefont {Koshibae}},
  \bibinfo {author} {\bibfnamefont {A.}~\bibnamefont {Kikkawa}}, \bibinfo
  {author} {\bibfnamefont {Y.}~\bibnamefont {Taguchi}}, \bibinfo {author}
  {\bibfnamefont {N.}~\bibnamefont {Nagaosa}}, \bibinfo {author} {\bibfnamefont
  {Y.}~\bibnamefont {Tokura}},\ and\ \bibinfo {author} {\bibfnamefont
  {F.}~\bibnamefont {Kagawa}},\ }\bibfield  {title} {\bibinfo {title}
  {Nonthermal current-induced transition from skyrmion lattice to
  nontopological magnetic phase in spatially confined {MnSi}},\ }\href
  {https://doi.org/10.1103/PhysRevB.106.144425} {\bibfield  {journal} {\bibinfo
   {journal} {Physical Review B}\ }\textbf {\bibinfo {volume} {106}},\ \bibinfo
  {pages} {144425} (\bibinfo {year} {2022})}\BibitemShut {NoStop}%
\bibitem [{\citenamefont {Iwasaki}\ \emph
  {et~al.}(2013{\natexlab{a}})\citenamefont {Iwasaki}, \citenamefont
  {Mochizuki},\ and\ \citenamefont {Nagaosa}}]{Iwasaki2013}%
  \BibitemOpen
  \bibfield  {author} {\bibinfo {author} {\bibfnamefont {J.}~\bibnamefont
  {Iwasaki}}, \bibinfo {author} {\bibfnamefont {M.}~\bibnamefont {Mochizuki}},\
  and\ \bibinfo {author} {\bibfnamefont {N.}~\bibnamefont {Nagaosa}},\
  }\bibfield  {title} {\bibinfo {title} {Universal current-velocity relation of
  skyrmion motion in chiral magnets},\ }\href
  {https://doi.org/10.1038/ncomms2442} {\bibfield  {journal} {\bibinfo
  {journal} {Nature Communications}\ }\textbf {\bibinfo {volume} {4}},\
  \bibinfo {pages} {1463} (\bibinfo {year} {2013}{\natexlab{a}})}\BibitemShut
  {NoStop}%
\bibitem [{\citenamefont {Iwasaki}\ \emph
  {et~al.}(2013{\natexlab{b}})\citenamefont {Iwasaki}, \citenamefont
  {Mochizuki},\ and\ \citenamefont {Nagaosa}}]{Iwasaki2013_2}%
  \BibitemOpen
  \bibfield  {author} {\bibinfo {author} {\bibfnamefont {J.}~\bibnamefont
  {Iwasaki}}, \bibinfo {author} {\bibfnamefont {M.}~\bibnamefont {Mochizuki}},\
  and\ \bibinfo {author} {\bibfnamefont {N.}~\bibnamefont {Nagaosa}},\
  }\bibfield  {title} {\bibinfo {title} {Current-induced skyrmion dynamics in
  constricted geometries},\ }\href {https://doi.org/10.1038/nnano.2013.176}
  {\bibfield  {journal} {\bibinfo  {journal} {Nature Nanotechnology}\ }\textbf
  {\bibinfo {volume} {8}},\ \bibinfo {pages} {742} (\bibinfo {year}
  {2013}{\natexlab{b}})}\BibitemShut {NoStop}%
\bibitem [{\citenamefont {Thiele}(1973)}]{Thiele1973}%
  \BibitemOpen
  \bibfield  {author} {\bibinfo {author} {\bibfnamefont {A.~A.}\ \bibnamefont
  {Thiele}},\ }\bibfield  {title} {\bibinfo {title} {Steady-state motion of
  magnetic domains},\ }\href {https://doi.org/10.1103/PhysRevLett.30.230}
  {\bibfield  {journal} {\bibinfo  {journal} {Physical Review Letters}\
  }\textbf {\bibinfo {volume} {30}},\ \bibinfo {pages} {230} (\bibinfo {year}
  {1973})}\BibitemShut {NoStop}%
\bibitem [{\citenamefont {Sch\"utte}\ \emph {et~al.}(2014)\citenamefont
  {Sch\"utte}, \citenamefont {Iwasaki}, \citenamefont {Rosch},\ and\
  \citenamefont {Nagaosa}}]{Chutte2014}%
  \BibitemOpen
  \bibfield  {author} {\bibinfo {author} {\bibfnamefont {C.}~\bibnamefont
  {Sch\"utte}}, \bibinfo {author} {\bibfnamefont {J.}~\bibnamefont {Iwasaki}},
  \bibinfo {author} {\bibfnamefont {A.}~\bibnamefont {Rosch}},\ and\ \bibinfo
  {author} {\bibfnamefont {N.}~\bibnamefont {Nagaosa}},\ }\bibfield  {title}
  {\bibinfo {title} {Inertia, diffusion, and dynamics of a driven skyrmion},\
  }\href {https://doi.org/10.1103/PhysRevB.90.174434} {\bibfield  {journal}
  {\bibinfo  {journal} {Physical Review B}\ }\textbf {\bibinfo {volume} {90}},\
  \bibinfo {pages} {174434} (\bibinfo {year} {2014})}\BibitemShut {NoStop}%
\bibitem [{\citenamefont {Song}\ \emph {et~al.}(2024)\citenamefont {Song},
  \citenamefont {Wang}, \citenamefont {Zhang}, \citenamefont {Liu},
  \citenamefont {Wang}, \citenamefont {Zheng}, \citenamefont {Tian},
  \citenamefont {Dunin-Borkowski}, \citenamefont {Zang},\ and\ \citenamefont
  {Du}}]{Song2024}%
  \BibitemOpen
  \bibfield  {author} {\bibinfo {author} {\bibfnamefont {D.}~\bibnamefont
  {Song}}, \bibinfo {author} {\bibfnamefont {W.}~\bibnamefont {Wang}}, \bibinfo
  {author} {\bibfnamefont {S.}~\bibnamefont {Zhang}}, \bibinfo {author}
  {\bibfnamefont {Y.}~\bibnamefont {Liu}}, \bibinfo {author} {\bibfnamefont
  {N.}~\bibnamefont {Wang}}, \bibinfo {author} {\bibfnamefont {F.}~\bibnamefont
  {Zheng}}, \bibinfo {author} {\bibfnamefont {M.}~\bibnamefont {Tian}},
  \bibinfo {author} {\bibfnamefont {R.~E.}\ \bibnamefont {Dunin-Borkowski}},
  \bibinfo {author} {\bibfnamefont {J.}~\bibnamefont {Zang}},\ and\ \bibinfo
  {author} {\bibfnamefont {H.}~\bibnamefont {Du}},\ }\bibfield  {title}
  {\bibinfo {title} {Steady motion of 80-nm-size skyrmions in a 100-nm-wide
  track},\ }\href {https://doi.org/10.1038/s41467-024-49976-6} {\bibfield
  {journal} {\bibinfo  {journal} {Nature Communications}\ }\textbf {\bibinfo
  {volume} {15}},\ \bibinfo {pages} {5614} (\bibinfo {year}
  {2024})}\BibitemShut {NoStop}%
\bibitem [{\citenamefont {Tatara}\ and\ \citenamefont
  {Fukuyama}(2014)}]{Tatara2014}%
  \BibitemOpen
  \bibfield  {author} {\bibinfo {author} {\bibfnamefont {G.}~\bibnamefont
  {Tatara}}\ and\ \bibinfo {author} {\bibfnamefont {H.}~\bibnamefont
  {Fukuyama}},\ }\bibfield  {title} {\bibinfo {title} {Phasons and excitations
  in skyrmion lattice},\ }\href {https://doi.org/10.7566/JPSJ.83.104711}
  {\bibfield  {journal} {\bibinfo  {journal} {Journal of the Physical Society
  of Japan}\ }\textbf {\bibinfo {volume} {83}},\ \bibinfo {pages} {104711}
  (\bibinfo {year} {2014})}\BibitemShut {NoStop}%
\bibitem [{\citenamefont {Hoshino}\ and\ \citenamefont
  {Nagaosa}(2018)}]{Hoshino2018}%
  \BibitemOpen
  \bibfield  {author} {\bibinfo {author} {\bibfnamefont {S.}~\bibnamefont
  {Hoshino}}\ and\ \bibinfo {author} {\bibfnamefont {N.}~\bibnamefont
  {Nagaosa}},\ }\bibfield  {title} {\bibinfo {title} {Theory of the magnetic
  skyrmion glass},\ }\href {https://doi.org/10.1103/PhysRevB.97.024413}
  {\bibfield  {journal} {\bibinfo  {journal} {Physical Review B}\ }\textbf
  {\bibinfo {volume} {97}},\ \bibinfo {pages} {024413} (\bibinfo {year}
  {2018})}\BibitemShut {NoStop}%
\bibitem [{\citenamefont {Ishikawa}\ \emph {et~al.}(1976)\citenamefont
  {Ishikawa}, \citenamefont {Tajima}, \citenamefont {Bloch},\ and\
  \citenamefont {Roth}}]{Ishikawa1976}%
  \BibitemOpen
  \bibfield  {author} {\bibinfo {author} {\bibfnamefont {Y.}~\bibnamefont
  {Ishikawa}}, \bibinfo {author} {\bibfnamefont {K.}~\bibnamefont {Tajima}},
  \bibinfo {author} {\bibfnamefont {D.}~\bibnamefont {Bloch}},\ and\ \bibinfo
  {author} {\bibfnamefont {M.}~\bibnamefont {Roth}},\ }\bibfield  {title}
  {\bibinfo {title} {Helical spin structure in manganese silicide {MnSi}},\
  }\href {https://doi.org/https://doi.org/10.1016/0038-1098(76)90057-0}
  {\bibfield  {journal} {\bibinfo  {journal} {Solid State Communications}\
  }\textbf {\bibinfo {volume} {19}},\ \bibinfo {pages} {525} (\bibinfo {year}
  {1976})}\BibitemShut {NoStop}%
\bibitem [{\citenamefont {Anderson}\ and\ \citenamefont
  {Kim}(1964)}]{Anderson1964}%
  \BibitemOpen
  \bibfield  {author} {\bibinfo {author} {\bibfnamefont {P.~W.}\ \bibnamefont
  {Anderson}}\ and\ \bibinfo {author} {\bibfnamefont {Y.~B.}\ \bibnamefont
  {Kim}},\ }\bibfield  {title} {\bibinfo {title} {Hard superconductivity:
  Theory of the motion of abrikosov flux lines},\ }\href
  {https://doi.org/10.1103/RevModPhys.36.39} {\bibfield  {journal} {\bibinfo
  {journal} {Reviews of Modern Physics}\ }\textbf {\bibinfo {volume} {36}},\
  \bibinfo {pages} {39} (\bibinfo {year} {1964})}\BibitemShut {NoStop}%
\bibitem [{\citenamefont {Bauer}\ and\ \citenamefont
  {Pfleiderer}(2012)}]{Bauer2012}%
  \BibitemOpen
  \bibfield  {author} {\bibinfo {author} {\bibfnamefont {A.}~\bibnamefont
  {Bauer}}\ and\ \bibinfo {author} {\bibfnamefont {C.}~\bibnamefont
  {Pfleiderer}},\ }\bibfield  {title} {\bibinfo {title} {Magnetic phase diagram
  of {MnSi} inferred from magnetization and {AC} susceptibility},\ }\href
  {https://doi.org/10.1103/PhysRevB.85.214418} {\bibfield  {journal} {\bibinfo
  {journal} {Physical Review B}\ }\textbf {\bibinfo {volume} {85}},\ \bibinfo
  {pages} {214418} (\bibinfo {year} {2012})}\BibitemShut {NoStop}%
\end{thebibliography}%

\end{document}


\title{Emergent reactance induced by the deformation of a current-driven skyrmion lattice: Supplementary Information}

\author{Matthew T. Littlehales}
\email{matthew.t.littlehales@durham.ac.uk}
\affiliation{Durham University, Department of Physics, South Road, Durham, DH1 3LE, United Kingdom}
\affiliation{ISIS Neutron and Muon Source, Rutherford Appleton Laboratory, Didcot OX11 0QX, United Kingdom}

\author{Max T. Birch}
\affiliation{RIKEN Center for Emergent Matter Science (CEMS), Wako, Japan}

\author{Akiko Kikkawa}
\affiliation{RIKEN Center for Emergent Matter Science (CEMS), Wako, Japan}

\author{Yasujiro Taguchi}
\affiliation{RIKEN Center for Emergent Matter Science (CEMS), Wako, Japan}

\author{Diego Alba Venero}
\affiliation{ISIS Neutron and Muon Source, Rutherford Appleton Laboratory, Didcot OX11 0QX, United Kingdom}

\author{Peter D. Hatton}
\affiliation{Durham University, Department of Physics, South Road, Durham, DH1 3LE, United Kingdom}

\author{Naoto Nagaosa}
\affiliation{RIKEN Center for Emergent Matter Science (CEMS), Wako, Japan}
\affiliation{Fundamental Quantum Science Program, TRIP Headquarters, RIKEN, Wako 351-0198, Japan}

\author{Yoshinori Tokura}
\affiliation{RIKEN Center for Emergent Matter Science (CEMS), Wako, Japan}
\affiliation{Department of Applied Physics, University of Tokyo, Tokyo, Japan}
\affiliation{Tokyo College, University of Tokyo, Tokyo, Japan}

\author{Tomoyuki Yokouchi}
\email{tomoyuki.yokouchi@riken.jp}
\affiliation{RIKEN Center for Emergent Matter Science (CEMS), Wako, Japan}

\date{\today}

\maketitle

\setcounter{figure}{0}
\renewcommand{\figurename}{Fig.}
\renewcommand{\thefigure}{S\arabic{figure}}

\newpage
\section{Correction for Trivial Phase Rotation of Complex Impedance}
\begin{figure}[b!]
	\centering
	\includegraphics{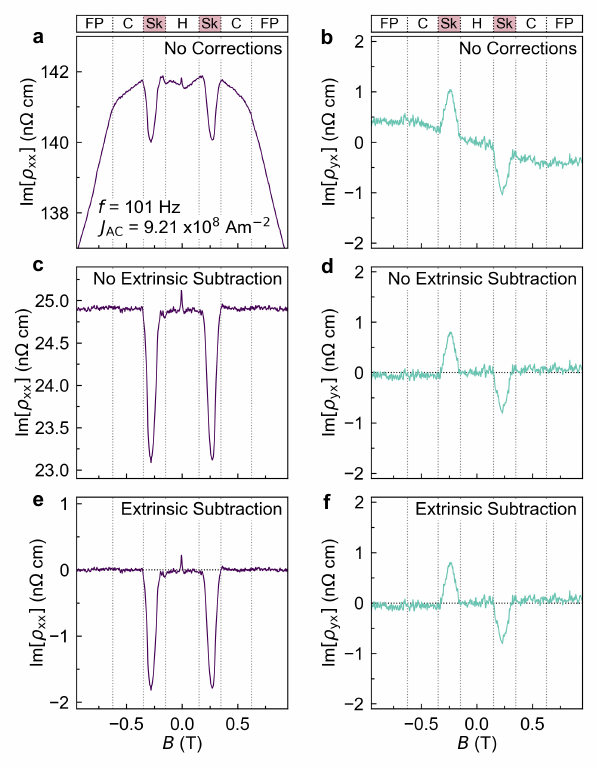}
	\caption{\textbf{Correction Procedure for Trivial Phase Rotation:} \textbf{a,b} Magnetic field dependence of the m[$\rho_\mathrm{xx}$] (\textbf{a}) and Im[$\rho_\mathrm{yx}$] (\textbf{b}) measured at $f$ = \SI{101}{\hertz} and $J_\mathrm{AC}$ = \SI{9.21}{\ampere\per\meter\squared}, uncorrected for trivial rotation. \textbf{c,d} Rotation of phase $\phi$ to bring the field polarised regime flat for both channels respectively, followed be a subtraction of extrinsic reactance in \textbf{e,f}.}
	\label{TrivialRotation}
\end{figure}



The impedance measurements include the both the resistance ($R$), inductance/capacitance ($X_\mathrm{Intrinsic}$) of our sample, and an extrinsic inductance/capacitance ($X_\mathrm{Extrinsic}$) arising from parasitic effects from the cables, connectors, and instruments, causing a systematic but frequency dependent rotation of the signal phase \cite{Choi2024}. In addition, any time-delay between the reference signal and the measured voltage results in mixing of the true real and imaginary components of the complex impedance as shown for both Im[$\rho_\mathrm{xx}$] and Im[$\rho_\mathrm{yx}$] in Fig. \ref{TrivialRotation}(a) and (b) respectively. We correct this effect by converting the measured real and complex values into a complex valued exponential with an additional variable phase according to the equation
\begin{equation*}
	Z = e^{-i\phi}\left[R + X_\mathrm{Intrinsic} + X_\mathrm{Extrinsic}\right],
\end{equation*}
where $\phi$ is the trivial rotation. We expect there to be no imaginary contributions to the resistivity for both Hall and longitudinal channels in the field polarised phase. Therefore, using a least-squares method, we first rotate the phase $\phi$ such that the field polarised regime is flat. The extrinsic reactance is symmetric with respect to the applied magnetic field and so is automatically removed for Im[$\rho_\mathrm{yx}$]. Whereas, for Im[$\rho_\mathrm{yx}$], we simply subtract this component to reveal the reactance from our sample. This process is outlined for each signal in Fig. \ref{TrivialRotation}. Note that all trivial phase rotation corrections are applied before any further analysis of the data. The remaining structures at zero magnetic field in Im[$\rho_\mathrm{xx}$] in Fig. \ref{TrivialRotation} (c,e) are due to intrinsic impedance from the PPMS cables. These signals have been measured on all test samples and are removed in the Hall channel in the antisymmetrisation process but cannot be removed in the longitudinal channel.


\newpage
\section{Estimation of topological Hall resistivity}
Our results in the main text rely on small changes to the topological Hall effect (THE) according to varying current density and frequency. For all measurements, we first subtract the ordinary Hall effect (OHE) by subtracting a straight line fit to the field polarised phase as demonstrated in Fig. \ref{SI_Subtractions}(a). Next, two methods were used in estimating the THE; for varying temperatures, and constant temperatures. These methods are outlined below.

\subsection*{Varying temperature measurements: } Due to the differing phase boundaries and anomalous Hall effect (AHE) with changing temperature, we first estimate the THE contribution through a linear fitting process. Fig. \ref{SI_Subtractions}(b) outlines the process in which we subtract fit a straight line to the conical regime. This analysis assumes that the anomalous Hall effect is proportional to the magnetic field. Strictly speaking, the anomalous Hall resistivity is proportional to the magnetization. However, since the magnetization is nearly linearly proportional to the magnetic field around the skrymion phase in MnSi \cite{Bauer2012}, this procedure provides a good approximation, serving as a useful method to estimate the THE when the magnetic phase boundaries vary with temperature, but leads to discrepancies arising in the helical phase from a differing gradient, as demonstrated in Fig. \ref{SI_Subtractions}(c). This method was used to estimate the magnitude and extent of the THE contribution from the SkL in Fig. 1e in the main text.

\subsection*{Constant temperature measurements:} 
\begin{figure*}[!]
    \centering
    \includegraphics[width = \textwidth]{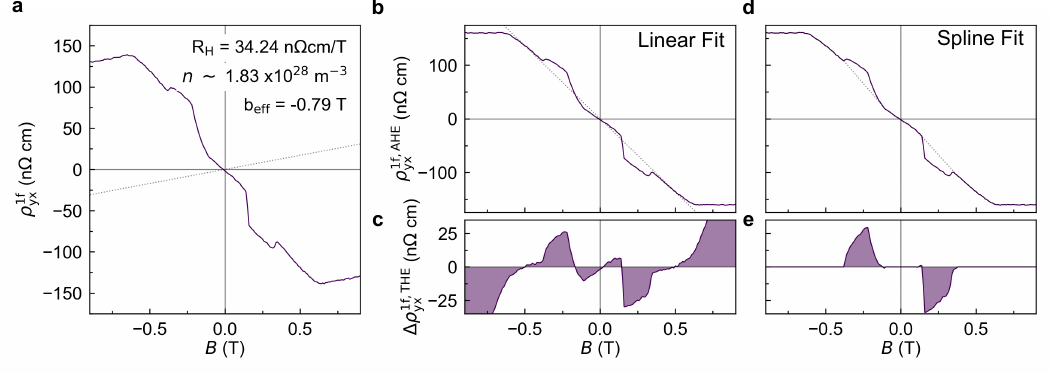}
    \caption{\textbf{Ordinary Hall effect (OHE) measurements and anomalous Hall effect (AHE) subtractions.} The Hall resistivity $\rho_\mathrm{yx}$ is plotted as a function of magnetic field $B$ in (\textbf{a}) measured at 23 K using $J_\mathrm{AC} = $\SI{1.74e7}{\ampere\per\meter\squared}. Above saturation ($\sim$ \SI{0.7}{\tesla}), we assume that the signal is dominated by the ordinary Hall effect and fit a straight line to acquire the Hall coefficient $R_\mathrm{H}$. (b,c) Subtraction following the process outlined for varying temperature. We subtract a straight line fit to the conical phase leading to discrepancies in the THE effect in the helical and field polarised regimes. (d,e) Spline fit assuming constant phase boundaries and linear interpolation through SkL phase. Less discrepancies are found in the THE (e) and using the THE at the center of the skyrmion phase $\rho_\mathrm{yx}^\mathrm{THE}(B = $ \SI{0.24}{\tesla}) = \SI{-27.06}{\nano\ohm\cm}, we estimate the effective magnetic field to be $b_\mathrm{eff} = $ \SI{-0.79}{\tesla}.} 
    \label{SI_Subtractions}
\end{figure*}

For the constant temperature measurements presented in Figures. 2-4, it is important to determine accurate changes in the THE. Consequently, we assume that both the SkL phase boundaries in magnetic field remained constant for varying current density and frequency, and that the AHE scales linearly through the SkL phase. Accordingly, we obtained the subtraction by first fitting a spline curve to the data that is linearly interpolated through the SkL phase and then subtracting this spline fit from the raw data. An example of this subtraction is presented in Fig.  \ref{SI_Subtractions}(d), with the result of the subtraction shown in Fig. \ref{SI_Subtractions}(e). By definition, the helical, conical and field polarised phases therefore exhibit no THE contribution, and this method provides a more accurate determination of small changes to the THE. No appreciable differences to the interpretation of the main results are found with the two differing methods.

\newpage
\section{Emergent Reactance measurements using DC Bias Current}
\begin{figure*}[!]
    \centering
    \includegraphics{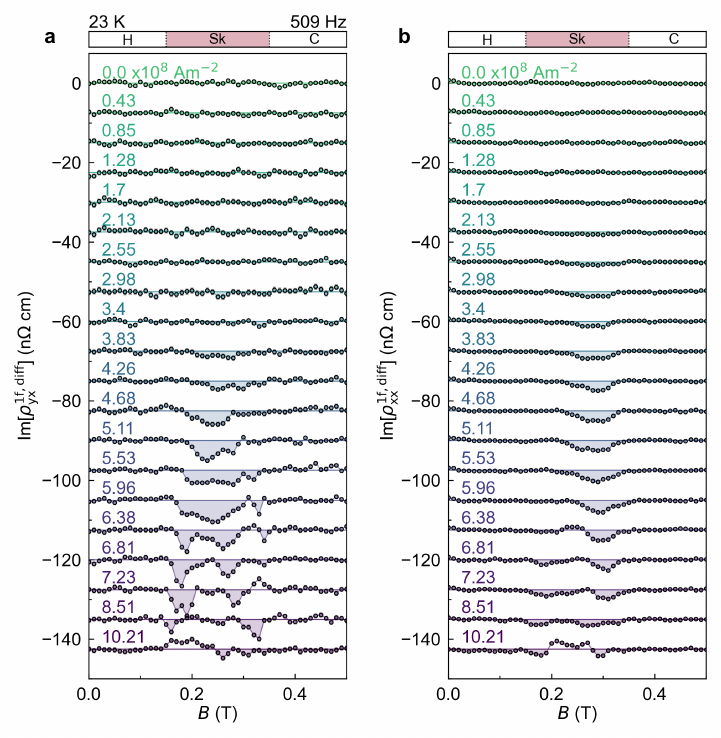}
    \caption{\textbf{The imaginary part of the complex impedance in DC bias measurements:} Field dependence of Im[$\rho_\mathrm{yx}^\mathrm{1f,diff}$] (\textbf{a}) and Im[$\rho_\mathrm{xx}^\mathrm{1f,diff}$] (\textbf{b}) for increasing $J_\mathrm{DC}$. The line plots are offset for clarity.}
    \label{ExtraDC_raw}
\end{figure*}

\begin{figure*}[t]
    \centering
    \includegraphics[width = \textwidth]{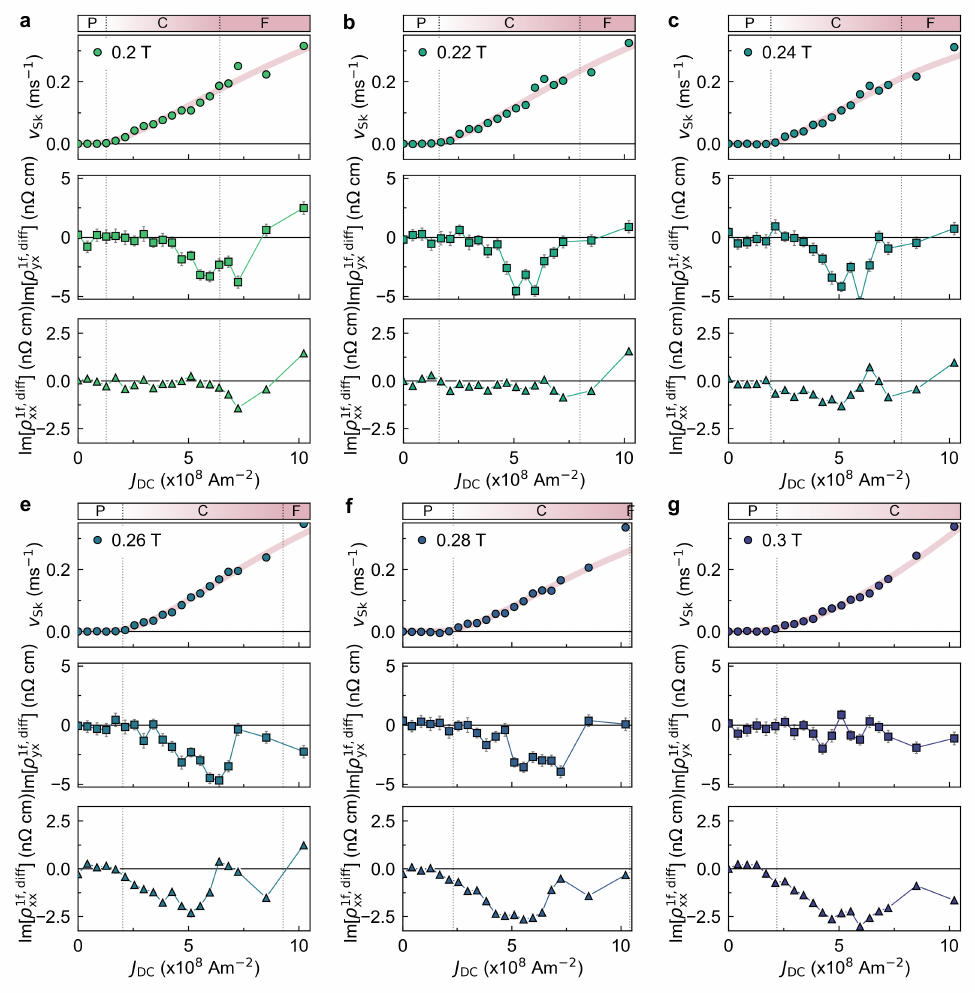}
    \caption{\textbf{Comparison of $J_\mathrm{DC}$ dependence for selected magnetic fields} $J_\mathrm{DC}$ dependencies for 
    $v_\mathrm{Sk}$ (top subplot), Im[$\rho_\mathrm{yx}^\mathrm{1f,diff}$] (middle subplot), and Im[$\rho_\mathrm{xx}^\mathrm{1f,diff}$] (bottom subplot). Qualitatively the behavior is independent of $B$.}
    \label{ExtraDC_flow}
\end{figure*}

The reactance signal is also observed in reactance measurements using a DC bias current. In Fig. \ref{ExtraDC_raw} we show the raw data of the imaginary component of the differential complex impedance Im[$\rho_\mathrm{yx}^\mathrm{1f,diff}$] (Fig. \ref{ExtraDC_raw}a) and Im[$\rho_\mathrm{xx}^\mathrm{1f,diff}$] (Fig. \ref{ExtraDC_raw}b), as a function of the magnetic field for various DC bias current $J_\mathrm{DC}$. It is clear for all  $J_\mathrm{DC}$ that the signal is constrained to the SkL phase. In Fig. \ref{ExtraDC_flow}, we show $\Delta$Re[$\rho_\mathrm{yx}^\mathrm{1f,diff}$], Im[$\rho_\mathrm{yx}^\mathrm{1f,diff}$], and Im[$\rho_\mathrm{xx}^\mathrm{1f,diff}$]  for selected magnetic fields within the SkL phase. Qualitatively, the behavior is independent of the magnetic field, while the creep and flow thresholds vary slightly with $B$. The imaginary components in both the transverse and longitudinal complex impedance reach their maximum in the creep region, followed by a reduction in the flow region, consistent with the AC measurements shown in Fig. 3 in the main text. This highlights the notion that SkL dynamics in the creep region is responsible for the reactance \cite{Birch2024}.


\newpage
\section{Scaling Behavior}
\begin{figure}[!]
    \centering
    \includegraphics{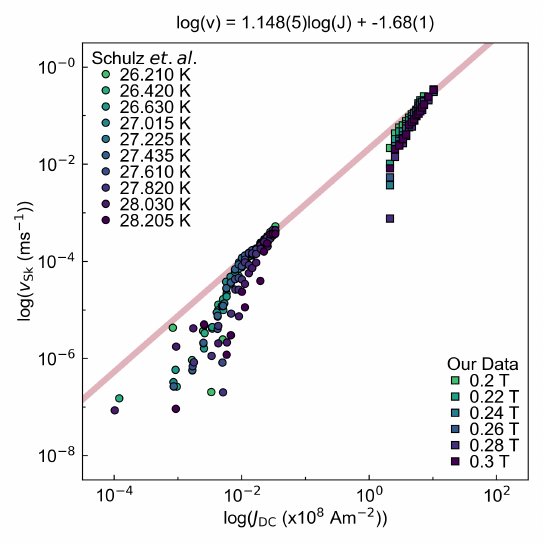}
    \caption{Scaling plot comparing the skyrmion velocity $J_\mathrm{DC}$ dependence of Schulz \textit{et. al.,} (circles) at different temperatures \cite{Schulz2012}, and our data (squares) for varying magnetic fields. }
    \label{Scaling}
\end{figure}

In Fig. \ref{Scaling}, we show the skyrmion velocity estimated from the reduction of the topological Hall resistivity as a function of the current density for bulk MnSi \cite{Schulz2012}  and our MnSi thin-plate device. While the critical current densities differ substantially, once the SkL enters the flow regime, for both datasets the skyrmion velocity follows the scaling according to $v_\mathrm{Sk} \propto J^{1.15}$ which agrees with the theoretical expectation of $v_\mathrm{Sk} \propto J$. This result indicates that the dynamical properties of our device are similar to those of MnSi single crystal, apart from the increase in the critical current densities. The higher critical current densities can be attributed to enhanced collective pinning, caused by an increased number of pinning sites introduced during the FIB fabrication process and confinement effects in  microfabricated samples \cite{Iwasaki2013, Iwasaki2013_2, Birch2024}.

\newpage
\section{Phenomenological calculation of emergent reactance in skyrmion lattice}

\subsection*{Transverse reactance induced by inertial motion of skyrmion lattice}
The velocity of the translational motion of the skyrmion $v_\mathrm{sk}$ obeys the Thiele equation: 
\begin{equation}
    m_\mathrm{sk} \dot{\bm{v}}_{\rm{sk}}+\mathcal{G} \times({\bm{v}}_{\mathrm{e}}-\bm{v}_\mathrm{sk}) +  \mathcal{D} ({\beta\bm{v}}_{\rm{e}} -\alpha\bm{v}_\mathrm{sk}) - \bm{\nabla}V_\mathrm{pin} = 0,
\end{equation}
where $\mathcal{G}$,  $\mathcal{D}$, $\beta$, $\alpha$, $V_\mathrm{pin}$, and $m_\mathrm{sk}$ are the gyro-coupling vector, dissipative force tensor, dimensionless constant characterizing the nonadiabatic electron spin dynamics, Gilbert damping constant, pinning potential, and the skyrmion mass, respectively. Here, the skyrmion mass arises from a renormalization of the skyrmion deformation into a mass-like term in the Thiele equation. In the creep region, the pinning potential is anharmonic and exhibits multiple minima. Consequently, in the creep region, the time derivative of  $\bm{v}_\mathrm{Sk}$ and the nonlinear term $\bm{\nabla} V_\mathrm{pin}$ in the Thiele equation cause the phase of $\bm{v}_\mathrm{Sk}$ to shift relative to that of the input AC current $\bar{\bm{J}}_\mathrm{AC}\sin(\omega t)$.  Phenomenologically, the skyrmion velocity can be described by  $\bm{v}_\mathrm{sk} = \bm{v}'_\mathrm{sk}\sin(\omega t)+\bm{v}''_\mathrm{sk}\cos(\omega t)$. In this case, the emergent electric field induced by the translational motion of SkL is given by $\bm{e}_\mathrm{em} = -\bm{v}_\mathrm{Sk}\times\bm{b}_\mathrm{em} = -\left[ \bm{v}'_\mathrm{sk}\sin(\omega t)+\bm{v}''_\mathrm{sk}\cos(\omega t)\right]\times\bm{b}_\mathrm{em}$, which gives the transverse reactance component as $\mathrm{Im}\rho_{yx} = Pb_\mathrm{em}v''_\mathrm{sk}/j_\mathrm{AC}$. 

In contrast, in the flow region, since $m_\mathrm{Sk}=0$ and $\bm{\nabla} V_\mathrm{pin}\sim 0 $  due to the absence of the internal deformation and a relative reduction of pinning force, the skyrmion velocity is equal to the electron velocity. Therefore, the emergent electric field induced by the translational motion of SkL is given by  $\bm{e}_\mathrm{em} = -\bm{v}_\mathrm{Sk}\times\bm{b}_\mathrm{em} = -\bm{v}_\mathrm{e}\times\bm{b}_\mathrm{em} \propto \bm{j}_\mathrm{AC}\sin(\omega t)\times\bm{b}_\mathrm{em} $, and the transverse reactance component disappears.

\subsection*{Longitudinal emergent reactance induced by deformation of skyrmion lattice }
Here, we discuss the relationship between the emergent electric field and phason and spin-tilting modes of SKL.
The magnetic moment of the SkL can be described as a superposition of three helices including its deformation as follows: 
\begin{align}
    \bm{m} &= M_z\bm{\hat{z}} + \sum_{i = a,b,c}\bm{m}_{i},\\
    \bm{m}_i &= M_h(\beta_i\bm{\hat{Q}}_i + \sqrt{1-\beta_i^2}\bm{l}_i),\\
    \bm{l}_i &= \bm{\hat{z}}\cos(\bm{Q}_i\cdot\bm{r}+\varphi_i) + (\bm{\hat{Q}}_i\times\bm{\hat{z}})\sin(\bm{Q}_i\cdot\bm{r} + \varphi_i),
\end{align}
where $\bm{Q}_i$ and $\bm{\hat{Q}}_i$ are the wavevector and it's unit vector for each helix. The three $\bm{\hat{Q}}_i$ vectors have a relative angle of 120 degrees.  Here,  $\varphi_i$ and $\beta_i$ are the phason and spin-tilting mode for each helix, respectively, which characterize the SkL deformation \cite{Tatara2014}. Generally, both  $\varphi_i$ and $\beta_i$ are functions of time and space.
As a simple example, we assume that one of the \textit{Q} vectors is parallel to the current direction ($x$ direction) and only the spin-tilting mode along the current direction is excited [i.e., $\beta_a = \beta_a(x,y,t)$ and $ \beta_b = \beta_c = \varphi_a = \varphi_b =\varphi_c = 0$]. The emergent electric fields along $x$ and $y$ directions are given by 
\begin{align}
e_x &= \frac{\hbar}{2e} \bm{n} \cdot (\partial_x \bm{n} \times \partial_t \bm{n})\\
e_y &= \frac{\hbar}{2e} \bm{n} \cdot (\partial_y \bm{n} \times \partial_t \bm{n}),
\end{align}
respectively. Here, $\bm{n}$ is the direction of the moment $\bm{n}= \bm{m}/|\bm{m}|$. Approximating $\beta_a \ll1 $ and $\bm{n}\approx \bm{m}$, we insert Eq. (3) into Eqs. (6) and (7) and obtain the following equatinos:
\begin{align*}
e_x & = \frac{Q \hbar}{16e} \left[\cos \left(\frac{Q x}{2}\right) \cos \left(\frac{1}{2} \sqrt{3} Q y\right) \right(4 M+3 \cos (Q x)+15\left)+8 M \cos (Q x)\right.\\
&\qquad\qquad\qquad	+2 \left. \cos ^3\left(\frac{Q x}{2}\right) \cos \left(\frac{1}{2} \sqrt{3} Q y\right)+4 \left(\cos \left(\sqrt{3} Q y\right)+3\right)\right]\partial_t \beta_1(x,y,t)\\
e_y & = \frac{\sqrt{3} Q \hbar}{4 e} \left[ 2 - M + \cos(Qx) \right] \sin\left(\frac{Qx}{2}\right) \sin\left(\frac{\sqrt{3} Q y}{2}\right) \partial_t \beta_1(x,y,t).
\end{align*}
Here, these electric fields have a spatially oscillating component and a spatially constant component. The observable electric fields are given by the  spatial average as follows:
\begin{align}
\left\langle e_x \right\rangle & = \frac{3 \hbar Q}{4e} \, \partial_t \beta_1(x,y,t) \\
\left\langle e_y \right\rangle & = 0
\end{align}
Therefore, when only the spin-tilting mode along the current direction is excited, the resulting emergent electric field appears only in the longitudinal direction and is proportional to the time derivative of the spin-tilting mode. This result is the same as in the case of the spin helix \cite{Nagaosa2019}, apart from a constant factor. We note that, in the actual case, other phason and spin-tilting modes are also excited. In particular, in the case of helices, the spin-tilting and phason modes are responsible for the positive and negative emergent reactance, respectively \cite{Kurebayashi2021, Nagaosa2019}. Since the sign of the emergent reactance in the SkL is negative, the phason mode of the SkL is expected to play an important role in the present case. However, further theoretical investigation remains a subject for future research. 

\section{Temperature dependence of emergent reactance}

\begin{figure}
	\centering
	\includegraphics[width = \textwidth]{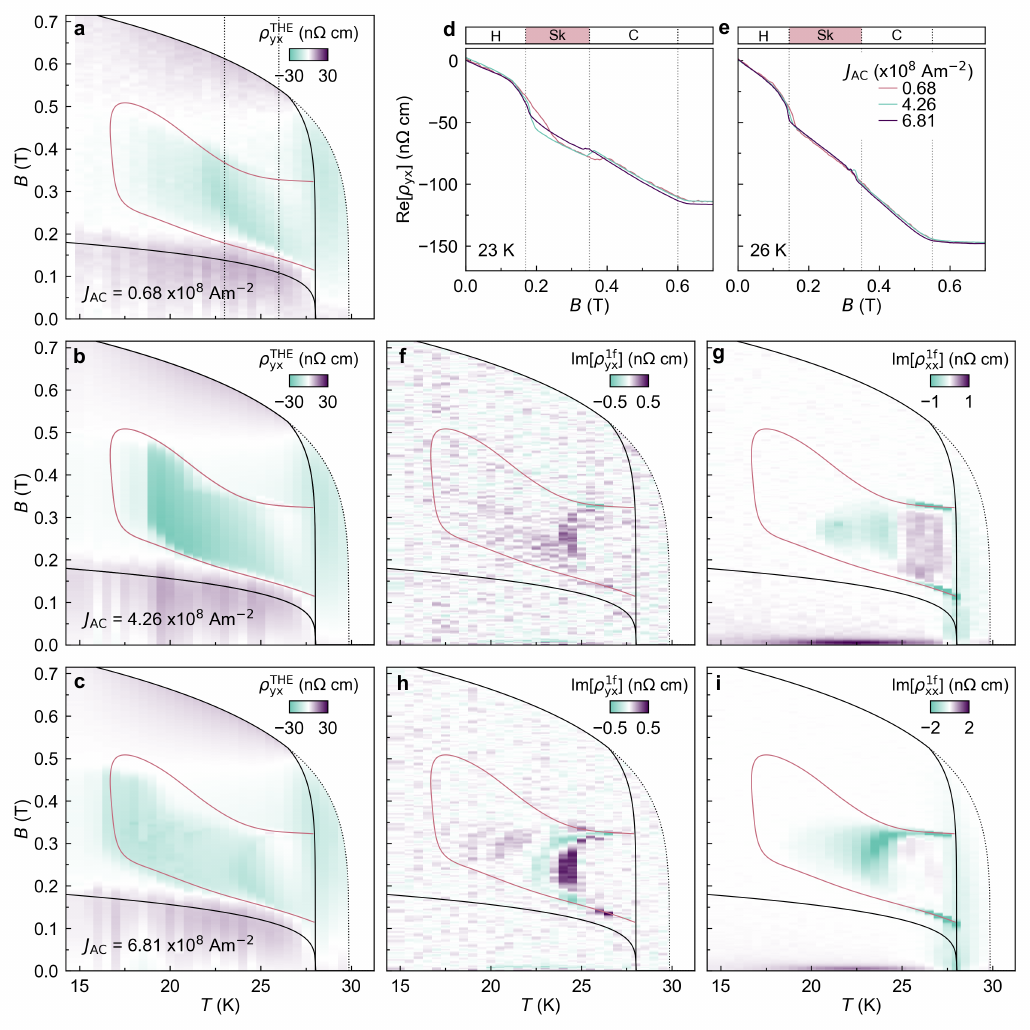}
	\caption{\textbf{Demonstration of SkL pinning close to the transition:} (a - c) Re[$\rho_\mathrm{yx}^\mathrm{THE}$] for differing current densities: $J_\mathrm{AC} =$ \SI{0.68e8}{\ampere\per\meter\squared}, \SI{4.26e8}{\ampere\per\meter\squared}, and \SI{6.81e8}{\ampere\per\meter\squared} respectively. (d,e) Magnetic field dependence of Re[$\rho_\mathrm{yx}^\mathrm{THE}$] for each of the three current densities at \SI{23}{\kelvin} (d) and \SI{26}{\kelvin} (e), clearly in (e) there is no change to the THE, indicating the skyrmions are pinned. (f,g) $B-T$ color maps of Im[$\rho_\mathrm{yx}^\mathrm{1f}$] and Im[$\rho_\mathrm{xx}^\mathrm{1f}$] measured with \SI{4.26e8}{\ampere\per\meter\squared}. Identical plots are shown in (h,i) for $J_\mathrm{AC}$ = \SI{6.81e8}{\ampere\per\meter\squared}. Notably, in (g), a positive Im[$\rho_\mathrm{xx}^\mathrm{1f}$] arises close to the fluctuation disordered regime. Since the SkL is pinned in this regime, we attribute this signal to the deformation of the pinned SkL \cite{Furuta2023}.}
	\label{Pinning}
\end{figure}

In Fig. \ref{Pinning}(a-c) we present the THE magnetic phase diagrams for three different current densities ($J_\mathrm{AC}$ = \SI{0.68e8}{\ampere\per\meter\squared}) (a), \SI{4.26e8}{\ampere\per\meter\squared} (b), and \SI{6.81e8}{\ampere\per\meter\squared} (c)). The THE is calculated using the temperature varying process outlined in section 1 of the supplementary information (hence the discrepancies in the helical and conical phases), and phase boundaries are overlaid according to those calculated for (b) using the method outlined in the methods section. The topological Hall effect exhibits distinct behaviour across three temperature ranges. As already shown in Fig. 2 of  the main text, in the centre of the skyrmion phase (22-25 K), the THE decreases with increasing current density. This trend is also confirmed by the isothermal cuts at 23 K measured at each current density (Fig. \ref{Pinning}d). The reduction of THE results from the current-induced motion of the SkL. In contrast, close to the paramagnetic transition temperature (25-28 K), THE does not depend on the current density as shown in Fig. \ref{Pinning}a-c and Fig. \ref{Pinning}e. This behaviour suggests that the SkL remains pinned. Finally, at low temperatures (below 22 K), the THE measured at low current density is smaller than that observed at the center of the SkL phase, indicating that the skyrmion density is reduced due to its coexistence with the conical phase. Furthermore, with increasing current density, the THE increases, which can be ascribed to small Joule heating effects, and/or current induced skyrmion nucleation \cite{Lemesh2018}.
	

These results have some important implications for the interpretation of the temperature dependence of the emergent reactance. In correspondence with the THE, the emergent reactance also exhibits distinct behaviour across the three temperature ranges. In the centre of the skyrmion phase (22-25 K), both Im[$\rho_\mathrm{yx}$] and Im[$\rho_\mathrm{xx}$] show faint signal signatures. This temperature range corresponds to the range in which the SkL can move, as discussed above, supporting the conclusion that the reactance originates from the SkL motion. In contrast, close to the paramagnetic transition temperature (25-28 K), Im[$\rho_\mathrm{yx}$] falls below the noise level. The SkL remains pinned under these conditions (as discussed above), resulting in a vanishing emergent electric field arising from the translational motion. Interestingly, in this temperature range, we observed a positive Im[$\rho_\mathrm{xx}$]. This positive Im[$\rho_\mathrm{xx}$] can be interpreted as an emergent reactance arising from the current-induced deformation of pinned skyrmions, predicted to exhibit a reactance with a positive sign \cite{Furuta2023}. Finally, at low temperatures (below 22 K) both Im[$\rho_\mathrm{yx}$] and Im[$\rho_\mathrm{xx}$] drop below the noise level. The absence of the reactance signals at low temperatures is a consequence of the skyrmion metastability. Since metastable skyrmions are easily destroyed or created by current application, the dynamical phase diagram differs from that of the skyrmion lattice state. In addition, the phason and spin-tilting modes are not well-defined for isolated metastable skyrmions. Therefore, only the equilibrium skyrmions lattice exhibits the expected dynamics discussed in the main text, and the emergent reactance.


\begin{figure}
	\centering
	\includegraphics{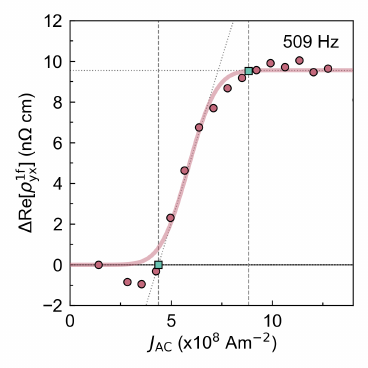}
	\caption{Example data of method used to disseminate creep and flow thresholds (square markers). Data fitting process is outlined in the Methods in the main text.}
	\label{JBoundaries}
\end{figure}

\newpage
\section{Joule Heating Considerations}
Here, we discuss the possible effects of Joule heating in our measurements and demonstrate that they cannot be responsible for the effects observed in this work.

\subsection*{Time-varying Joule Heating Effect}
\begin{figure}[b]
    \centering
    \includegraphics{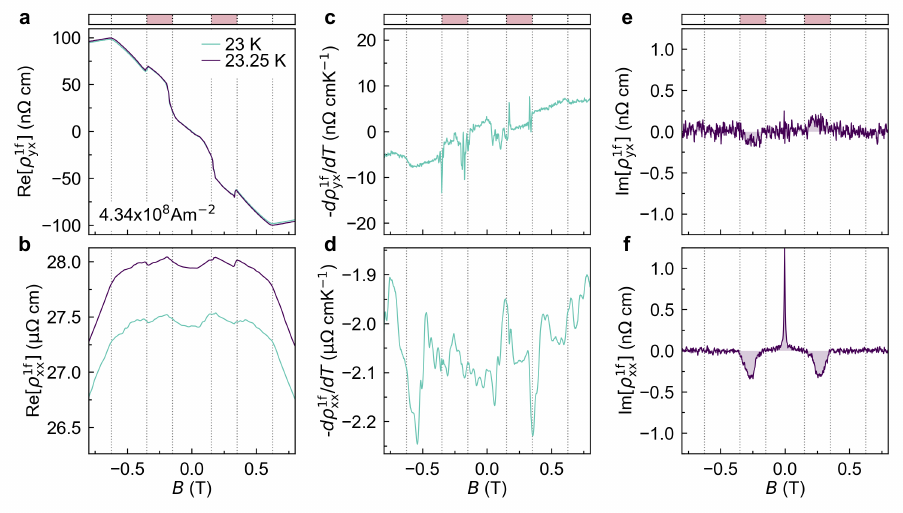}
    \caption{\textbf{Time-varying Joule Heating Effect}  \textbf{a,b} Magnetic field dependence of the Hall resistivity Re[$\rho_\mathrm{yx}^\mathrm{1f}$] (\textbf{a}) and magneto-resistivity Re[$\rho_\mathrm{xx}^\mathrm{1f}$] (\textbf{b}) measured using PPMS at heat bath temperatures of  \SI{23}{\kelvin} (teal), and \SI{23.25}{\kelvin} (purple). \textbf{c,d} $-\frac{d\rho_\mathrm{yx}^\mathrm{1f}}{dT}$ and $-\frac{d\rho_\mathrm{xx}^\mathrm{1f}}{dT}$calculated from the magnetic field dependences in (\textbf{a,b}), respectively. \textbf{e,f} Magnetic field dependence of Hall reactance Im[$\rho_\mathrm{yx}^\mathrm{1f}$] (\textbf{e}) and longitudinal reactance Im[$\rho_\mathrm{xx}^\mathrm{1f}$] (\textbf{f}). }
    \label{ACHeating}
\end{figure}
Theoretical models suggest that a time-varying resistance change due to the time-varying Joule heating effect may produce a non-linear reactance  \cite{Furata2024}. The main claim of this theory is that the imaginary part of the complex resistivity Im[$\rho^\mathrm{1f}$] arises from a delayed response of the time-varying resistance change as a consequence of the delay in the thermal relaxation process. In this model, Im[$\rho^\mathrm{1f}$] is proportional to the temperature derivative of the resistivity  i.e. $\frac{d\mathrm{Re}[\rho]}{dT}$. Thus, a useful test to determine whether time-varying Joule heating is responsible is to compare the magnetic-field dependence of  $\frac{d\mathrm{Re}[\rho]}{dT}$ and  Im[$\rho^\mathrm{1f}$]. We present $\frac{d\mathrm{Re}[\rho]}{dT}$ and  Im[$\rho^\mathrm{1f}$] for both the Hall and longitudinal resistivities in Fig. \ref{ACHeating} . Here, the temperature derivative of the resistivity is approximated by the difference in resistivity between 23 K and 23.25 K. It is clear that the magnetic-field dependence of Im[$\rho^\mathrm{1f}$] does not resemble that of $-\frac{d\rho}{dT}$. Therefore,  we conclude that time-varying Joule heating cannot account for the observed effects.


\subsection*{Estimation of Joule heating due to DC bias current }
\begin{figure}
    \centering
    \includegraphics{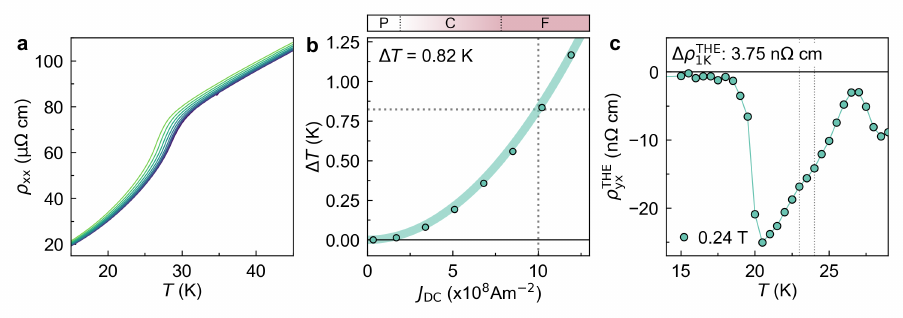}
    \caption{\textbf{Estimation of Joule heating due to DC bias current:} \textbf{a} longitudinal resistivity $\rho_\mathrm{xx}$ as a function of temperature for various DC bias currents.  \textbf{b} Temperature increase caused by the DC bias current when the heat bath of PPMS is set to $T = $ \SI{23}{\kelvin}. For $J_\mathrm{DC} = $\SI{10e8}{\ampere\per\meter\squared}, the temperature change is estimated to be $\Delta T = $ \SI{0.8}{\kelvin}. \textbf{c} Topological Hall resistivity as a function of versus temperature at $B = $ \SI{0.24}{\tesla}. The change in THE between \SI{23}{\kelvin} and \SI{24}{\kelvin} is approximately $\Delta\rho_\mathrm{yx}^\mathrm{THE} = $ \SI{3.75}{\nano\ohm\cm}.}
    \label{DCHeating}
\end{figure}
In the case of AC+DC measurements, the DC bias current $J_\mathrm{DC}$ may induce a constant temperature increase, which could reduce the topological Hall effect. To confirm that this temperature change cannot account for the observed reduction in THE, we estimate the temperature increase when the heat bath (i.e., the temperature of the sample holder) is set to \SI{23}{\kelvin}. We assume that the resistivity change under the application of $J_\mathrm{DC}$, relative to that measured at $J_\mathrm{DC} = $ \SI{0}{\ampere\per\meter\squared}, is caused by the temperature increase due to the Joule heating, and we calculate the corresponding temperature increases. In Fig. \ref{DCHeating}(b), we show the estimated temperature increase as a function of  $J_\mathrm{DC}$. Our estimation is consistent with a previous study of MnSi thin-plate device, which reports a temperature increase of 0.2-\SI{0.3}{\kelvin} for $J_\mathrm{DC}$ = \SI{7.7e8}{\ampere\per\meter\squared} \cite{Sato2022}. For $J_\mathrm{DC} = $ \SI{10e8}{\ampere\per\meter\squared}, the maximum current density in this study, the temperature increase is estimated to be $\Delta T\approx$\SI{0.8}{\kelvin}, which would lead to a change in THE of $\Delta\rho_\mathrm{yx}^\mathrm{THE}\approx$ \SI{3.77}{\nano\ohm\cm}, comparable to the noise level in the AC + DC measurement. Therefore,  the observed reduction in THE $\Delta\rho_\mathrm{yx}^\mathrm{THE}\approx$ \SI{20}{\nano\ohm\cm} cannot be attributed to the Joule heating effects. We thus conclude that the change in THE observed in our AC+DC measurements originates from the emergent electric field induced by the translational motion of the SkL.

\newpage
\bibliography{bib}